\documentclass[12pt]{iopart}

\newcommand{\be}{\begin{eqnarray}}
\newcommand{\ee}{\end{eqnarray}}
\newcommand{\bea}{\begin{eqnarray}}
\newcommand{\eea}{\end{eqnarray}}

\usepackage{graphicx}
\usepackage{bm}
\usepackage{color}
\usepackage{slashed}
\usepackage{cite}
\usepackage{graphicx}
\usepackage{multicol}
\usepackage{graphicx}
\usepackage{color}
\usepackage{slashed}
\usepackage{CJKutf8}
\begin{document}
\begin{CJK}{UTF8}{<font>}
\title{Observable characteristics of the charged black hole surrounded by thin disk accretion in Rastall gravity}

\author{Sen Guo$^{1}$, \ Guan-Ru Li$^{1}$, \ En-Wei Liang$^{*1}$}

\address{$^1$Guangxi Key Laboratory for Relativistic Astrophysics, School of Physical Science and Technology, Guangxi University, Nanning 530004, People's Republic of China}

\ead{sguophys@st.gxu.edu.cn; lew@gxu.edu.cn}
\vspace{10pt}
\begin{indented}
\item[]Apr 2022
\end{indented}

\begin{abstract}
The observable characteristics of the charged black hole (BH) surrounded by a thin disk accretion are investigated in the Rastall gravity. We found that the radii of the direct emission, lensing ring, and photon ring dramatically increased as the radiation field parameter increases, but they only weakly depend on the BH charge. Three positions of the radiation accretion disk relative to the BH are considered, i.e., the innermost accretion disk is closed to the radii of the innermost stable circular orbit, the photon ring of the BH, and the event horizon of the BH. The observed images in three cases respectively are obtained. It is found that the total observed flux is dominated by the direct emission, the lensing ring provides a small contribution, and the photon ring is negligible. The lensing and photon rings could not be observed in the blurred image with the EHT resolution. Our results suggest that the observable characteristics of the charged BH surrounded by the thin disk accretion in the Rastall gravity depend on both the BH space-time structure and the position of the radiating accretion disk with respect to the BH. The research of these BH images may serve as a probe for the BH-disk structure in M87$^{*}$ like nearby active galactic nuclei.
\end{abstract}


\section{Introduction}
\label{intro}
\par
Detection of gravitational wave emission from black hole (BH) mergers with the Laser-Interferometer Gravitational Wave-Observatory (LIGO) presents convincing existence of BHs in the universe \cite{1}. The Event Horizon Telescope (EHT) obtains the ultra-high angular resolution image of the accretion flow around the supermassive BH in M87$^{*}$, which shows a bright ring-shaped lump of radiation surrounding a central black region of an estimated 6.5 billion solar masses \cite{2,3,4,5,6,7}. The image illustrates that the BH space is illuminated by a magnetically arrested accretion disk \cite{Akiyama-1,Akiyama-2}. The observed dark area in the image is called the ``BH shadow'', and the ring-shaped lump of radiation is called the ``photon ring'' \cite{8}.

\par
An astrophysical BH is generally surrounded by a luminous accretion matter. Possible observational characteristics of the BHs shadows surrounded by various accretion matters were investigated for a long time. By considering a geometrically thin and optically thick standard accretion disk model, Luminet suggested that the emergence of the shadow and photon ring depend on the position and profile of the accretion \cite{9}. Assuming that a hot optically-thin accretion surrounds the supermassive BH in the center of our Galaxy, Falcke $et~al.$ proposed that the BH shadow is equivalent to the gravitational lensing effect \cite{10}. Cunha $et~al.$ studied the gravitational lensing effect and investigated the shadow feature of a spherical BH surrounded by a thin and heavy accretion disk. They found that an almost equatorial observer could observe different patches of the sky near the equatorial plane \cite{11}. It is under debate whether the BH shadow is solely determined by the space-time metric or affected by the accretion patterns. Narayan $et~al.$ investigated the shadows of the Schwarzschild BH under the newton/simple spherical accretion model. They presented that the location of the shadow edge is independent of the inner radius at which the accreting gas stops radiating, implying that the positions of the accretion do not impact the shadow size \cite{18}. However, Gralla $et~al.$ argued that the size of a Schwarzschild BH shadow depends on the details of accretion by considering an optically and geometrically thin accretion disk model \cite{19}.

\par
Analysing the BHs shadow images under the various gravity theories would be robust probes for the observable characteristics of the BHs. In the framework of the Einstein-{\AE}ther gravity theory, Zhu $et~al.$ argued that the presence of the {\ae}ther field could affect the shadow size of the charged/slowly rotating BHs \cite{12}. By investigating the shadows of the Gauss-Bonnet BH and the quintessence dark energy BH with the static/infalling spherical accretions, Zeng $et~al.$ found that the BH's optical appearance depends on the geometry and physical properties of the accretion, and its luminosity is a function of the impact parameter \cite{13,14}. Gan $et~al.$ studied the shadow and photon sphere of an asymptotically flat hairy BH in the Einstein-Maxwell-scalar gravity theory. They found that a photon has two unstable circular orbits in a particular parameter regime, corresponding to two bright concentric rings with different radii \cite{15}. In the framework of the clouds of strings and quintessence, He $et~al.$ showed that the brightness distribution of the photon sphere is a normal function of the string attenuation factor \cite{16}. Guo $et~al.$ found that the luminosities of both the shadows and rings of the Hayward BH are affected by the accretion flow property and the BH magnetic charge \cite{17}.

\par
One of the essential elements of general relativity (GR) is the covariant conservation of the energy-momentum tensor. Based on the GR, Rastall proposed a new modified theory, which aims to relax the condition of covariant energy-momentum conservation. In the Rastall theory, the covariant derivative of the energy-momentum tensor satisfies $\nabla^{\mu}T_{\mu \nu}=\lambda \nabla_{\nu} R$, where $\lambda$ is the proportional constant that tests the deviation with the Einstein gravity \cite{Ras}. The static spherically symmetric solutions of the compact objects in the Rastall gravity are proposed \cite{Oli,Mor}. Heydarzade $et~al.$ calculated the static spherically symmetric BH solutions with the anisotropic fluid field in the Rastall gravity. It is found that the field describing the characteristics of this solution generally can be dust field, radiation field, or dark energy component \cite{20}. By utilizing the Hamilton-Jacobi method, Ali $et~al.$ analyzed the tunneling radiation and Hawking temperature of the neutral regular BH in the Rastall gravity and derived the tunneling rate and BH temperature \cite{Ali-1}. In order to further explore the characteristics of the BH shadow under the Rastall gravity framework, \"{O}vg\"{u}n $et~al.$ studied the shadow cast of non-commutative BH in the Rastall gravity. They found that the visibility of the resulting shadow depends on the non-commutative parameter \cite{AA}. By investigating the shadow of a rotating BH surrounded by an anisotropic fluid field in the Rastall theory, Kumar $et~al.$ found the rotating Rastall BH leads to a shadow with smaller size than the Kerr BH for a given value of the rotational parameter \cite{Kumar}. Note that these analyses do not consider the scenario that the astrophysical BHs are usually surrounded by accretion matters.

\par
In our previous work, we investigated the shadow and photon sphere of a charged BH with a perfect fluid radiation field (PFRF) surrounded by the static/infalling spherical accretion within the framework of the Rastall gravity. It is found that the shadow luminosity of this BH with infalling spherical accretion is dimmer than that of the static spherical accretion, but the photon sphere luminosity is brighter than the static one \cite{21}. It is well known that the radiation received by our telescope comes not from the BHs themselves, but instead originates in the accretion disks which surround them. The radiation leaks through the disk, escapes from its surface, and travels along space curved by the strong gravity of the BH. In order to be closer to the real astrophysical environment, we attempt to investigate the observable characteristics of a charged BH surrounded by a thin disk accretion in the Rastall gravity in this paper. By considering the three innermost radii of the radiation accretion disk relative to the BH, we analyze the shadows and rings as well as the corresponding observational images and compare the theoretical two-dimensional image with the EHT image.

\par
Our paper is organized as follows. Section \ref{sec:2} briefly reviews the effective potential of our BH and discusses the rings classification. In section \ref{sec:3}, we present the images of the shadows and rings as well as the corresponding observation images based on the three radiation positions of the accretion disk. We draw the conclusions in section \ref{sec:4}.

\section{BH effective potential and light deflection in the Rastall gravity}
\label{sec:2}
\par
The spherical symmetric space-time metric is
\begin{equation}
\label{1-1}
{\rm d}s^{2}=-f(r){\rm d}t^{2}+\frac{1}{f(r)}{\rm d}r^{2}+r^{2}{\rm d}\theta^{2}+r^{2}\sin^{2}\theta {\rm d}\phi^{2}.
\end{equation}
Considering the Rastall theory, the Einstein-Rastall field equation is given by \cite{Ali-2}
\begin{equation}
\label{1-2}
R_{\rm \mu\nu}-\frac{1}{2}g_{\rm \mu \nu } R + \kappa \lambda g_{\rm \mu \nu } R = \kappa T_{\rm \mu \nu },
\end{equation}
where $\kappa$ is the Rastall gravitational coupling constant. By solving the Einstein-Rastall field equation, the BH metric surrounded by a perfect fluid in the Rastall gravity can be written as \cite{Prih}
\begin{eqnarray}
\label{1-3}
{\rm d}s^{2}=&&-\Bigg(1-\frac{2M}{r}+\frac{Q^{2}}{r^{2}}+\frac{N_{\rm s}}{r^{\rm \zeta}}\Bigg){\rm d}t^{2}
+\frac{{\rm d}r^{2}}{\Big(1-\frac{2M}{r}+\frac{Q^{2}}{r^{2}}+\frac{N_{\rm s}}{r^{\rm \zeta}}\Big)}\nonumber \\
&&+r^{2}{\rm d}\theta^{2}+r^{2}\sin^{2}\theta {\rm d}\phi^{2},
\end{eqnarray}
where $N_{\rm s}$ is the surrounding field structure parameter, and $\zeta$ is
\begin{equation}
\label{1-4}
\zeta=\frac{1+3\omega-6 \kappa \lambda (1+\omega)}{\Big(1-3 \kappa \lambda (1+\omega)\Big)}.
\end{equation}
For the case of the charged BH surrounded by a PFRF ($\omega=1/3$) in the Rastall gravity, the metric potential reads as \cite{20}
\begin{equation}
\label{2-2}
f(r)=1-\frac{2M}{r}+\frac{Q^{2}-N_{\rm r}}{r^{2}},
\end{equation}
where $M$ is the mass and $Q$ is the charge of the BH. $N_{\rm r}$ is the radiation field parameter, depending on the energy density of the radiation field \cite{20}. When the $N_{\rm r} \rightarrow 0$, the metric (\ref{2-2}) will degenerate into the Reissner-Nordstr\"{o}m (RN) BH. In general, $f(r)=0$ admits two roots $r_{\pm}$,
\begin{equation}
\label{2-3}
r_{\pm}=M\pm \sqrt{M^{2}-Q^{2}+N_{\rm r}},
\end{equation}
where the symbol $``\pm"$ denotes that the event horizon radius $r_{+}$ and the inner horizon radius $r_{-}$. Note that the parameters $Q$ and $N_{\rm r}$ could be re-scaled to be dimensionless, depending on their dimensions related with $M$. Assuming that $q=Q/M$ and $n=N_{\rm r} / M^{2}$, the parameter $q$ as a function of $n$ is shown in Fig.1. For $q<\sqrt{n+1}$, the solution describes a BH with a Cauchy horizon $r_{-}$ and an event horizon $r_{+}$ (the green region of Fig.1). The extremal BH is obtained with $q=\sqrt{n+1}$, whose causal structure is characterized by the unique horizon $r_{\rm ex}=r_{+}=r_{-}$ (the dark-blue line in Fig.1). The case of $q>\sqrt{n+1}$ corresponds to a naked singularity.
\begin{center}
\includegraphics[width=6.5cm,height=5cm]{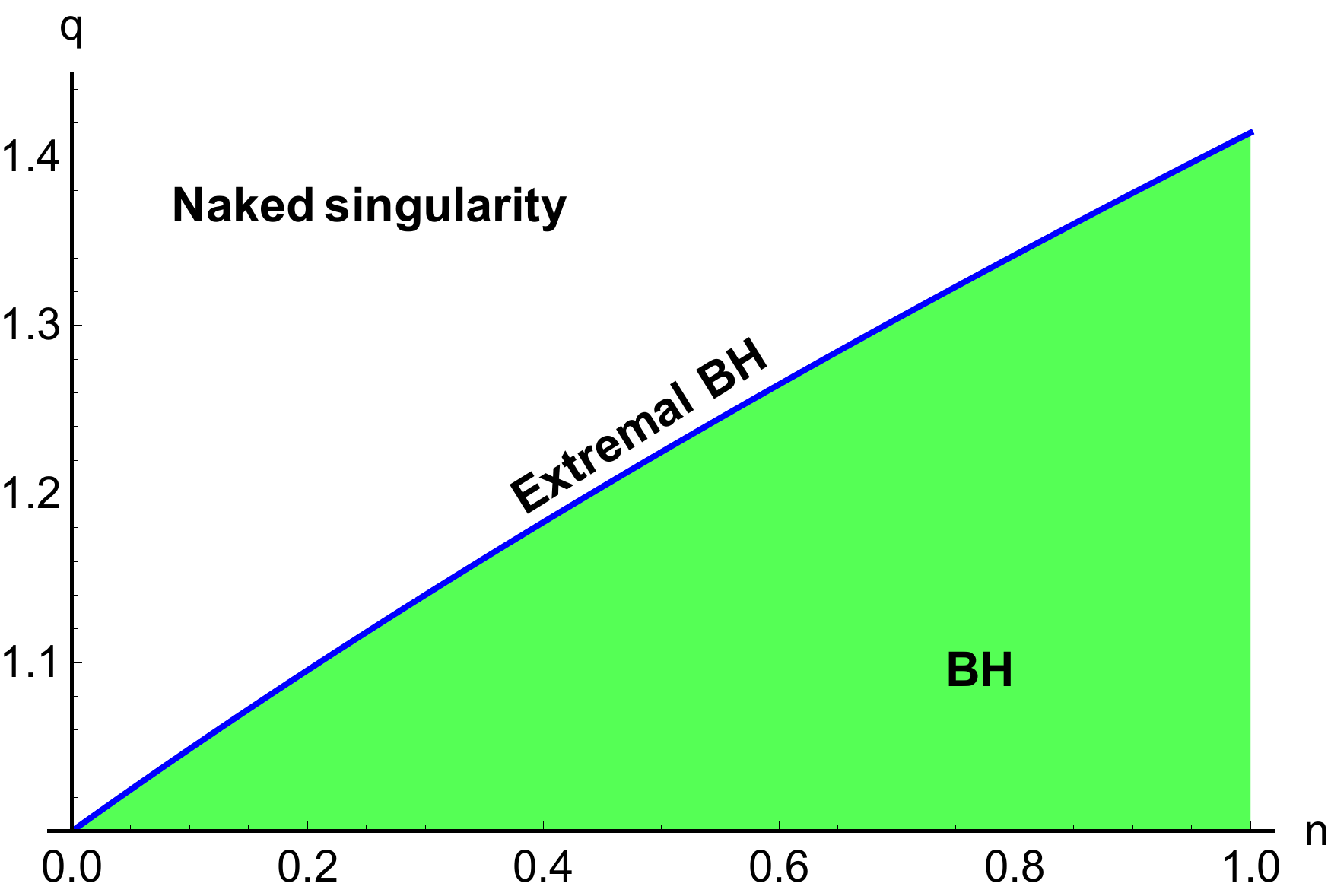}
\parbox[c]{15.0cm}{\footnotesize{\bf Fig~1.}  %. figure caption
The parameter $q$ as a function of $n$. The BH mass is taken as $M=1$.}
\label{fig1}
\end{center}

\par
The effective potential is
\begin{equation}
\label{2-4}
V_{\rm e}=\frac{f(r)}{r^{2}}=\frac{1}{r^{2}}\Bigg(1-\frac{2M}{r}+\frac{Q^{2}-N_{\rm r}}{r^{2}}\Bigg).
\end{equation}
The radius of the BH photon ring is derived from the following conditions
\begin{equation}
\label{2-5}
V_{\rm e}=\frac{1}{b_{\rm c}},~~~~V'_{\rm e}=0,
\end{equation}
where $b_{\rm c}$ is the critical impact parameter. Based on Eqs.(\ref{2-3})-(\ref{2-5}), the numerical results of the BH event horizon radii, shadow radii, and critical impact parameters with different parameters are reported in Tab.1 and Tab.2. One can observe that the increase of the BH charge leads to the decrease of their, displaying that the BH photon ring is shrunk inward the BH by increasing the charge. While the radiation field parameter have the opposite effect.
\begin{center}
\label{table:1}
{\footnotesize{\bf Table 1.} The event horizon radii, shadow radii and critical impact parameters for different $Q$ with $M=1$ and $N_{\rm r}=1$.\\
\vspace{2mm}
\begin{tabular}{ccccccc}
\hline
{Q} & {0} & {0.2} & {0.4} & {0.6} & {0.8} & {1}\\\hline
$r_{\rm +}$   & $2.42$ & $2.40$  &  $2.36$ &  $2.28$ &  $2.17$ &  $2.00$ \\
$r_{\rm ph}$  & $3.56$ & $3.54$ &  $3.48$ &  $3.38$ &  $3.22$ &  $3.00$ \\
$b_{\rm ph}$  & $5.94$ & $5.91$ &  $5.83$ &  $5.69$ &  $5.49$ &  $5.19$ \\
\hline
\end{tabular}}
\end{center}
\begin{center}
\label{table:2}
{\footnotesize{\bf Table 2.} The event horizon radii, shadow radii and critical impact parameters for different $N_{\rm r}$ with $M=1$ and $Q=0.5$.\\
\vspace{2mm}
\begin{tabular}{ccccccc}
\hline
{RF} & {0} & {1} & {2} & {3} & {4} & {5}\\\hline
$r_{\rm +}$   &  $1.87$ & $2.32$ & $2.66$ & $2.94$ &  $3.18$ &  $3.39$ \\
$r_{\rm ph}$  &  $2.82$ & $3.44$ & $3.89$ & $4.28$ &  $4.62$ &  $4.93$ \\
$b_{\rm ph}$  &  $4.97$ & $5.77$ & $6.39$ & $6.92$ &  $7.38$ &  $7.81$ \\
\hline
\end{tabular}}
\end{center}

\par
Note that the shadow radius depends on the BH charge, and can be measured with the EHT observations \cite{2,3,4,5,6,7}. Based on Eqs.(\ref{2-2}) and (\ref{2-5}), the left panel of Fig.2 shows the shadow radius of the BH in the Rastall gravity $r_{\rm sh}$ as a function of charge $Q$. It is found that $r_{\rm sh}$ decreases with the increase of $Q$. The shadow radius of the M87$^{*}$ is $r_{\rm M87^{*}}\approx 5.5 \pm 0.75r_{\rm g}$, where $r_{g}$ is the radius of the Schwarzschild BH \cite{Akiyama-1,Akiyama-2}. We constrain $Q$ with EHT observations. As illustrated in Fig.2, our result is consistent with that derived from the EHT observations within the observational uncertainty. Using the $1\sigma$ and $2\sigma$ confidence intervals of the $r_{\rm M87^{*}}$, the BH charge can be constrained as $Q \leq 1.03$ within $1 \sigma$ and $Q \leq 1.55$ within $2 \sigma$. In the same way, the radiation field parameter can be constrained as $N_{\rm r} \leq 3.23$ within $1 \sigma$ and $N_{\rm r} \leq 6.39$ within $2 \sigma$. We take a some values of $N_{\rm r}$ and $Q$ for illustrating our results in the later analysis.
\begin{center}
\includegraphics[width=6.5cm,height=5cm]{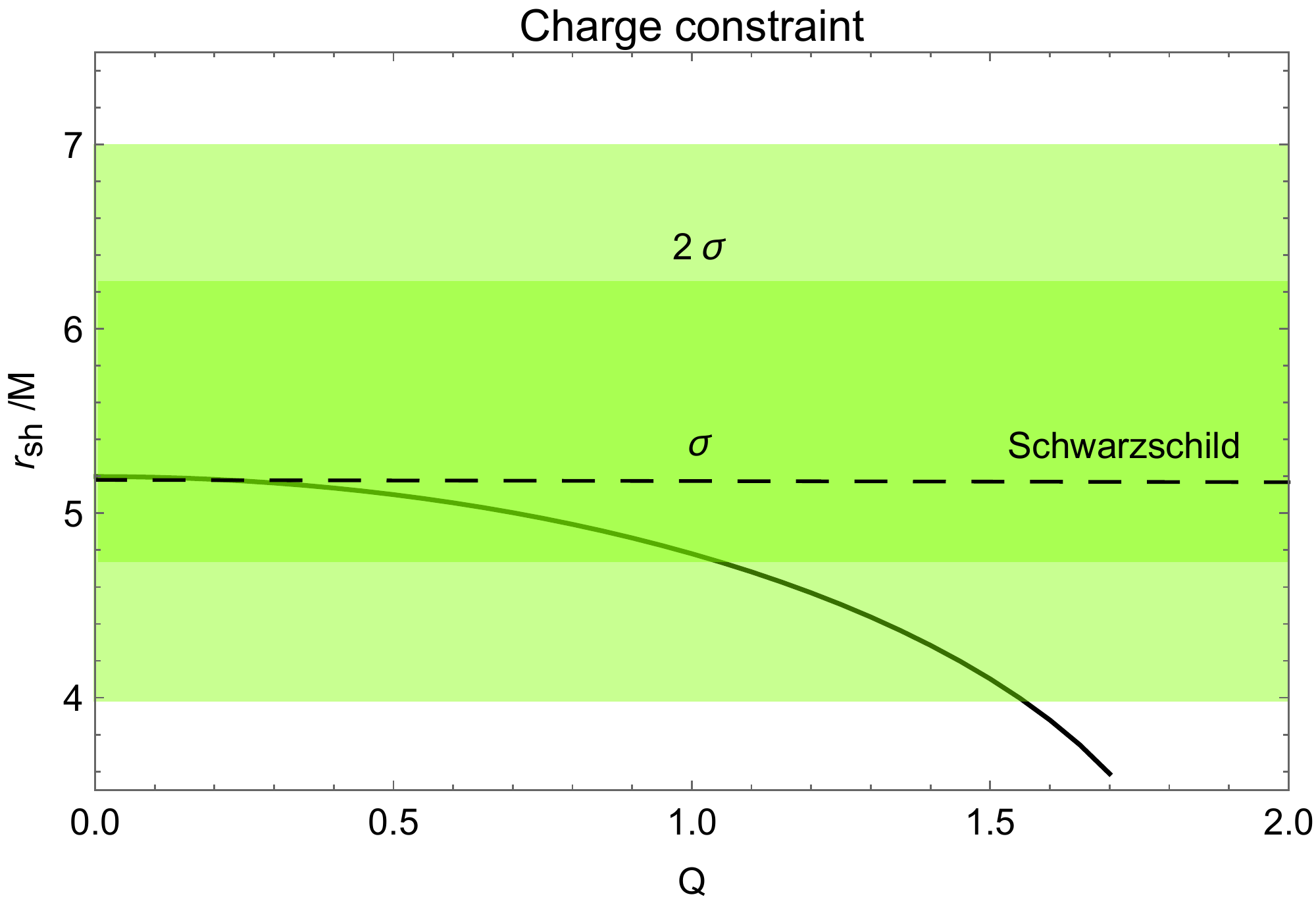}
\hspace{0.8cm}
\includegraphics[width=6.5cm,height=5cm]{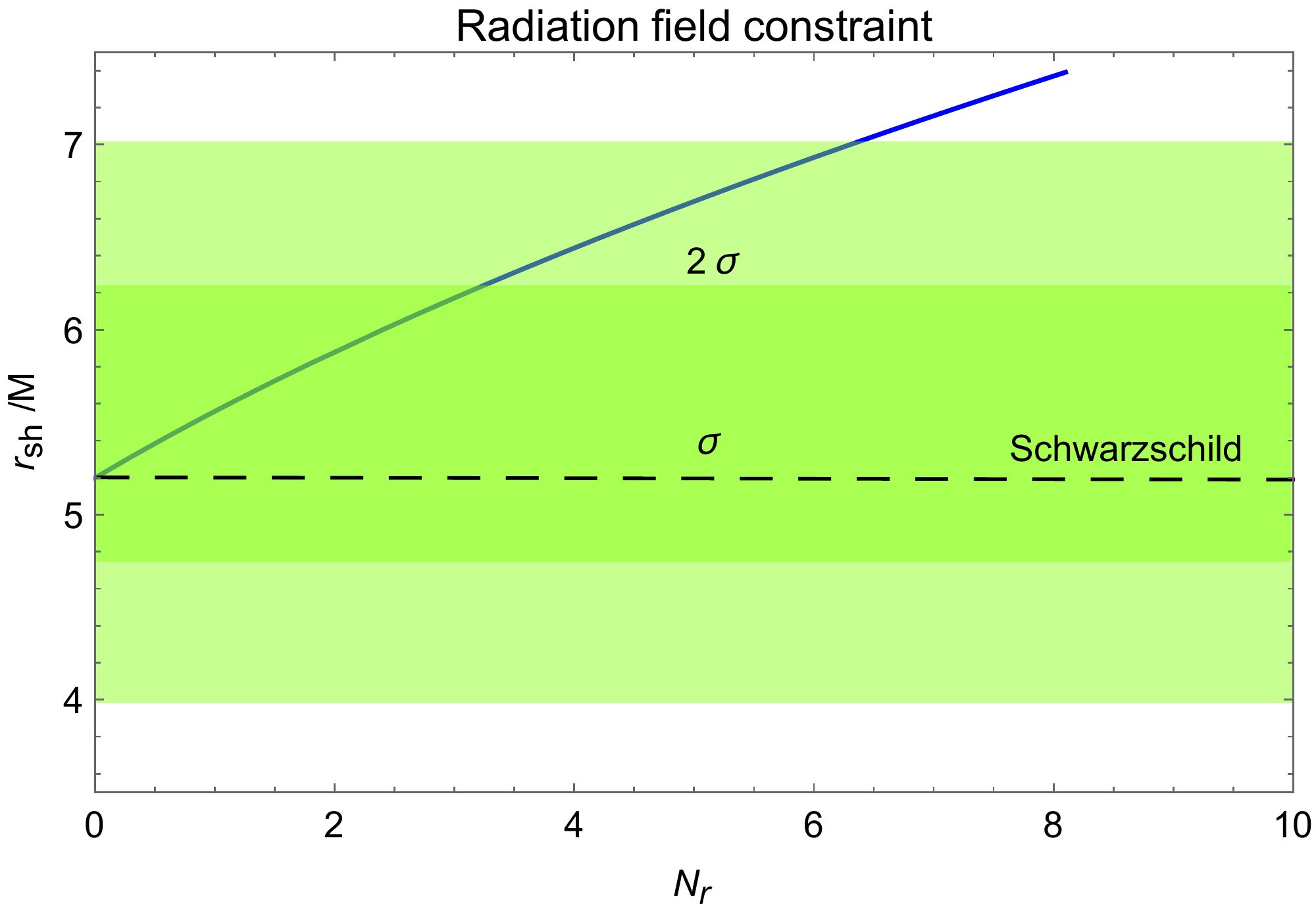}
\parbox[c]{15.0cm}{\footnotesize{\bf Fig~2.}  %. figure caption
Shadow radius as a function of BH charge $Q$ and radiation field parameter $N_{\rm r}$ for the charged BH with a PFRF in the Rastall theory, respectively. The yellow and green shaded regions refer to the areas that are $1\sigma$ and $2\sigma$ confidence levels with the M87$^{*}$ observations.}
\label{fig2}
\end{center}

\par
We assume that the radiation is from an optically and geometrically thin accretion disk in the equatorial plane of the BH. It emits isotropically in the rest frame of static worldlines, and a static observer at the north pole. Utilizing the ray-tracing method, the rings are seen by the observer can be subdivided into direct emission, lensing ring, and photon ring \cite{19}. The light of the direct emission falls on the front of the accretion disk, and it intersects the equatorial plane just once. For the lensing ring, the light breaks through the thin disk (the light trajectories intersect the equatorial plane twice) and falls on the back of the accretion disk. Thus, the light picks up additional brightness from the second intersection between the light and the accretion disk. The photon ring represents the light arriving at the front side of the accretion disk once again, leading to additional brightness from the third intersection. Therefore, the total observed intensity should be the sum of those intensities.

\par
Based on Liouville's theorem, $I_{\rm em}/({\upsilon_{\rm em}})^{3}$ is conserved in the direction of light propagation, where $I_{\rm em}$ is the radiation specific intensity of the accretion disk and $\upsilon_{\rm em}$ is the radiation frequency \cite{17}. For a charged BH with a PFRF surrounded by a thin accretion disk in the Rastall gravity, the integrated intensity can be obtained by integrating over the whole range of received frequencies,
\begin{eqnarray}
\label{2-6}
I_{\rm S}(r)=\int I_{\rm o}(r) {\rm d} \upsilon_{\rm o}=\Bigg(1-\frac{2M}{r}+\frac{Q^{2}-N_{\rm r}}{r^{2}}\Bigg)^{2} I_{\rm emit}(r),
\end{eqnarray}
where $I_{\rm o}(r)$ is the observed specific intensity with a single frequency $\upsilon_{\rm o}$, $I_{\rm emit}(r) \equiv \int I_{\rm em}(r) {\rm d} \upsilon_{\rm em}$ is defined as the total radiation intensity of the thin accretion disk. Hence, the total observed intensity ($\rm ergs^{-1}cm^{-2}str^{-1}Hz^{-1}$) by a static observer is
\begin{eqnarray}
\label{2-7}
I_{\rm obs}=\sum\limits_{n} \Bigg(1-\frac{2M}{r}+\frac{Q^{2}-N_{\rm r}}{r^{2}}\Bigg)^{2} I_{\rm emit}(r)|_{r=r_{\rm n}(b)},
\end{eqnarray}

\par
Figure 3 shows $r_{\rm n}(b)$ as a function of $b$ under several representative values of the parameters, where $r_{\rm n}(b)$ represents the radial coordinate of the $n_{\rm th}$ intersection between the light with impact parameter $b$ and the accretion disk. The slope of the $r_{\rm n}(b)$ - ${\rm d}r/{\rm d}b$ - is defined as the (de)magnification factor \cite{14}. For the case of $n=1$, the $r_{\rm n}(b)$ is a linear function with a slope approximately equal to 1, indicating that $r_{\rm n}$ is proportional to $b$ and it contributes the most of total observed flux. This corresponds to the ``direct emission'' scenario. The case of $n=2$ is for the ``lensing ring''. The $r_{\rm n}(b)$ function illustrates as an asymptotic curve, and $b$ is limited to a narrow range around $b \simeq 5.5M-6.3M$ ($Q=0.6, N_{\rm r}=1$). In this range of $b$, the observer will see a highly demagnified image of the back side of the disk, with a variable demagnification given by the slope of the curve. Over the displayed range of $r$, the average slope is around $62.5$, indicating that the lensing ring image is around $62.5$ times smaller, and hence will typically contribute around $1.6\%$ of the total flux. Therefore, the lensing ring should be a thin ring in the observation image. The case of $n=3$ is for the ``photon ring''. The $r_{\rm n}(b)$ is almost a vertical line in this scenario, suggesting that the photon ring should be an extremely thin ring. For the parameter value of $Q=0.6, N_{\rm r}=1$, the average slope is around $166$, its contribution to the total observed flux is only $\sim 0.6\%$. It is also find that the increase of the BH charge value leads to a slight decrease in the impact parameter. The radiation field parameter increases results in the increase of the impact parameter, which is consistent with that reported in Tab.1 and Tab.2.
\begin{center}
\includegraphics[width=6.5cm,height=4.7cm]{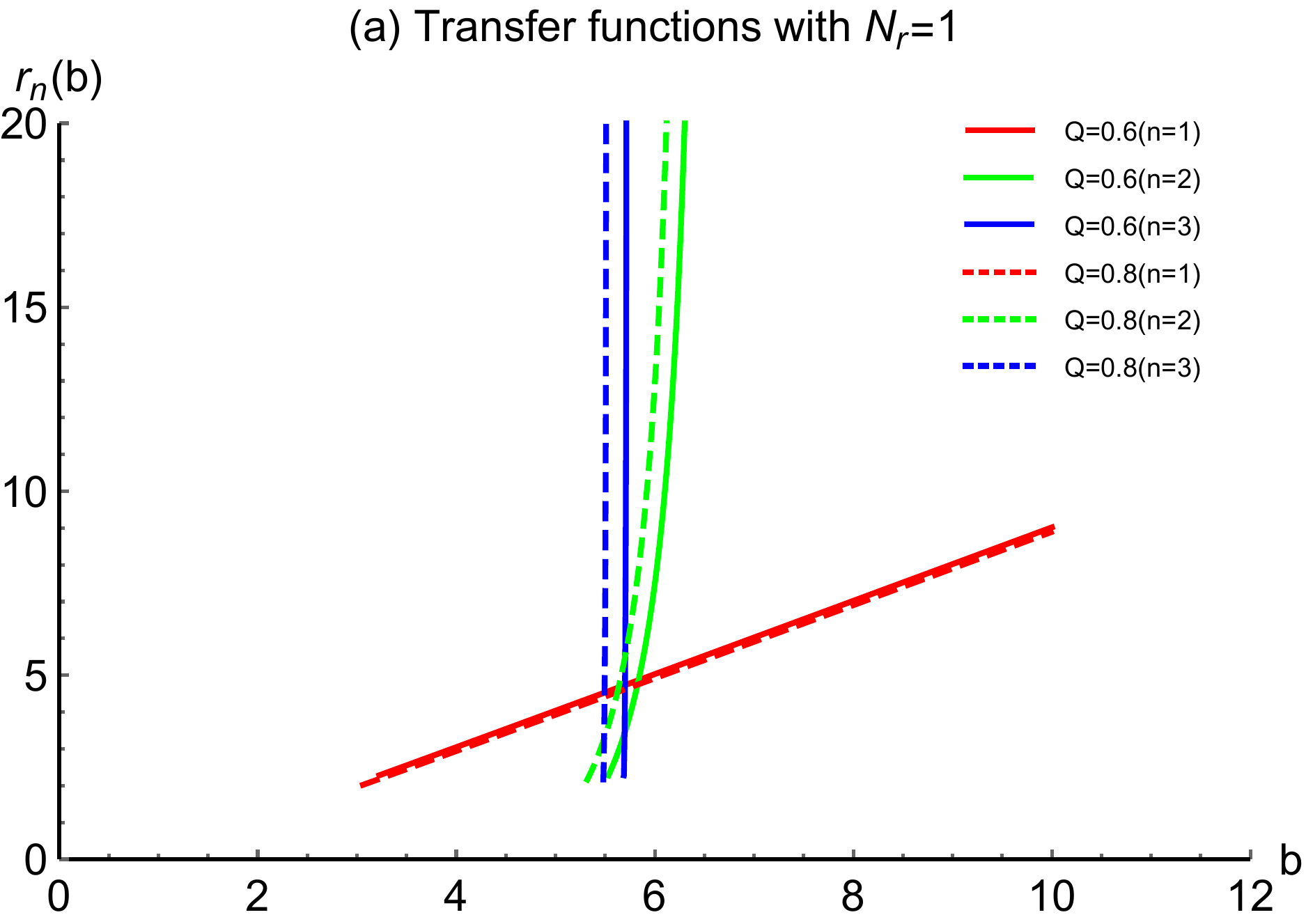}
\hspace{1.2cm}
\includegraphics[width=6.5cm,height=4.7cm]{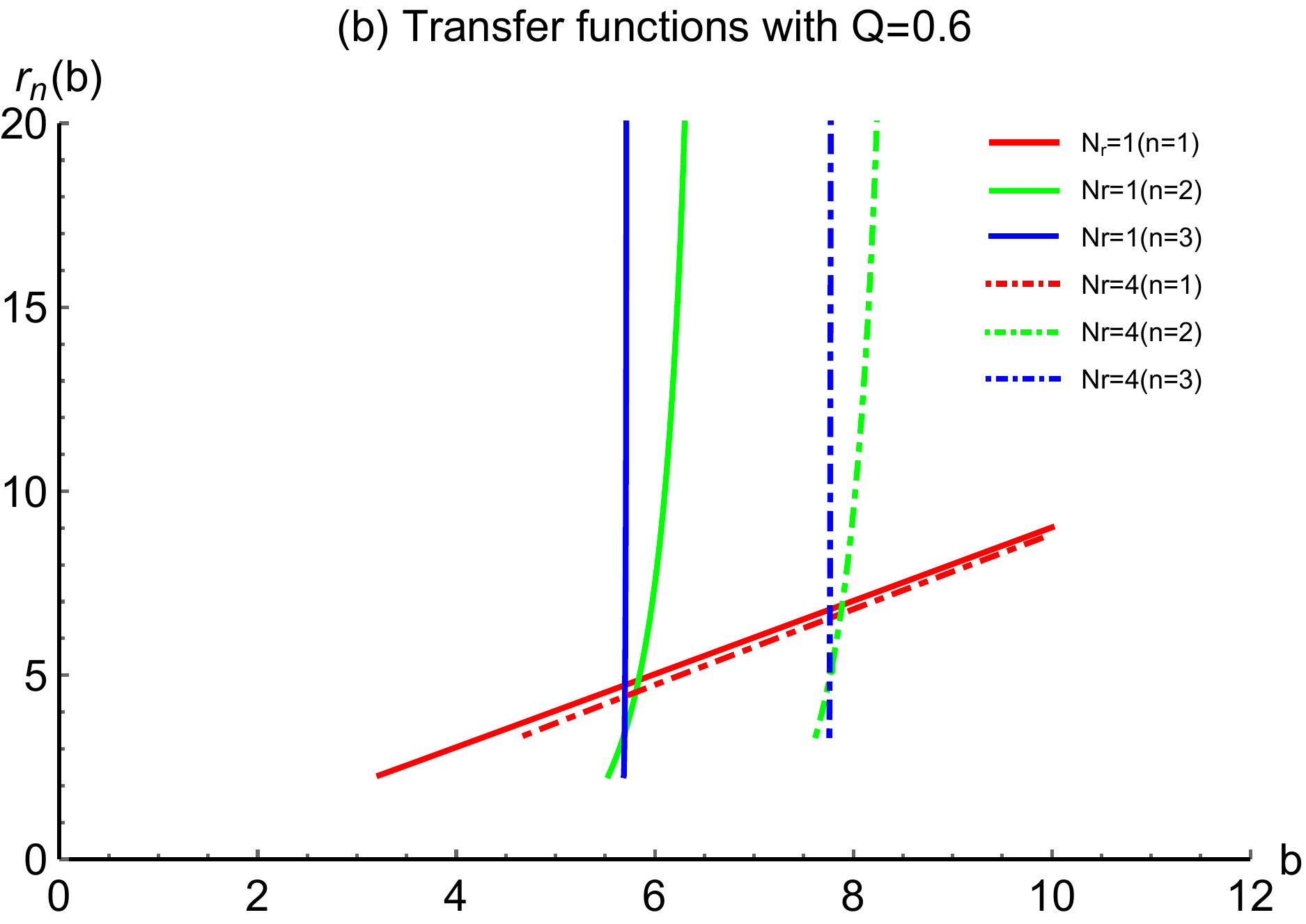}
%\hspace{0.1cm}
%\includegraphics[width=5cm,height=4.7cm]{tf-0.pdf}
\parbox[c]{15.0cm}{\footnotesize{\bf Fig~3.}  % figure caption
The $r_{\rm n}(b)$ as a function $b$ for different $Q$ and $N_{\rm r}$. The BH mass is taken as $M=1$.}
\label{fig3}
\end{center}

\par
Figure 4 illustrates the light trajectories of different rings in the polar coordinate by utilizing the ray-tracing code. One can see that the radius of the black disk is smaller for a larger charge, and the light rays are more curved near the BH since the increase of the BH charge leads to the increase of the space-time curvature. It is also found that the radii of the direct emission, lensing ring, and photon ring are slightly shrunk if the charge value increases. In addition to the BH charge, the radiation field parameter $N_{\rm r}$ can also affects the light trajectories of the charged BH in the Rastall gravity. It is found that that the radii of the direct emission, lensing ring, and photon ring dramatically increases as the radiation field parameter increases.
\begin{center}
\includegraphics[width=4.5cm,height=4.5cm]{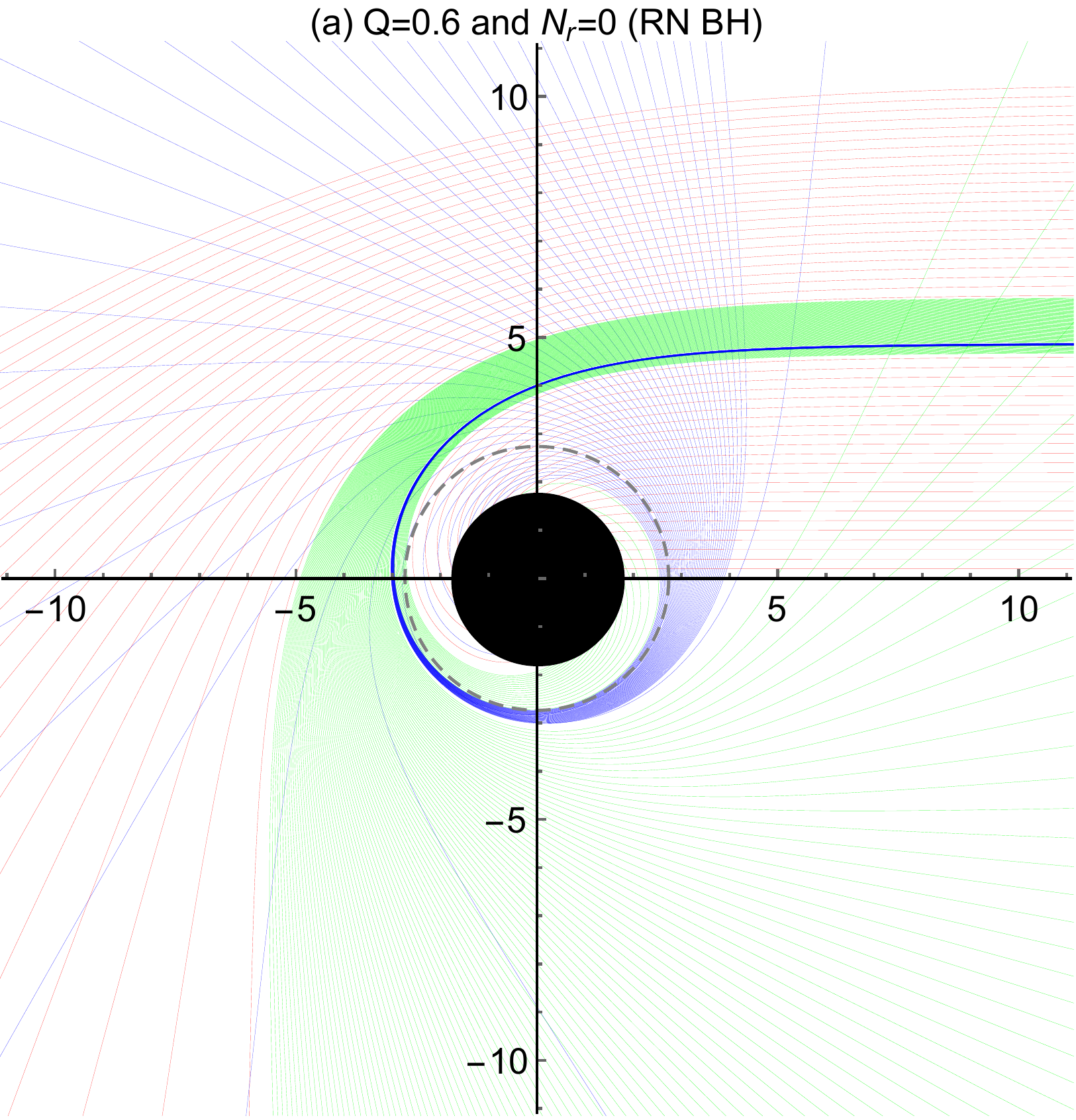}
\hspace{0.2cm}
\includegraphics[width=4.5cm,height=4.5cm]{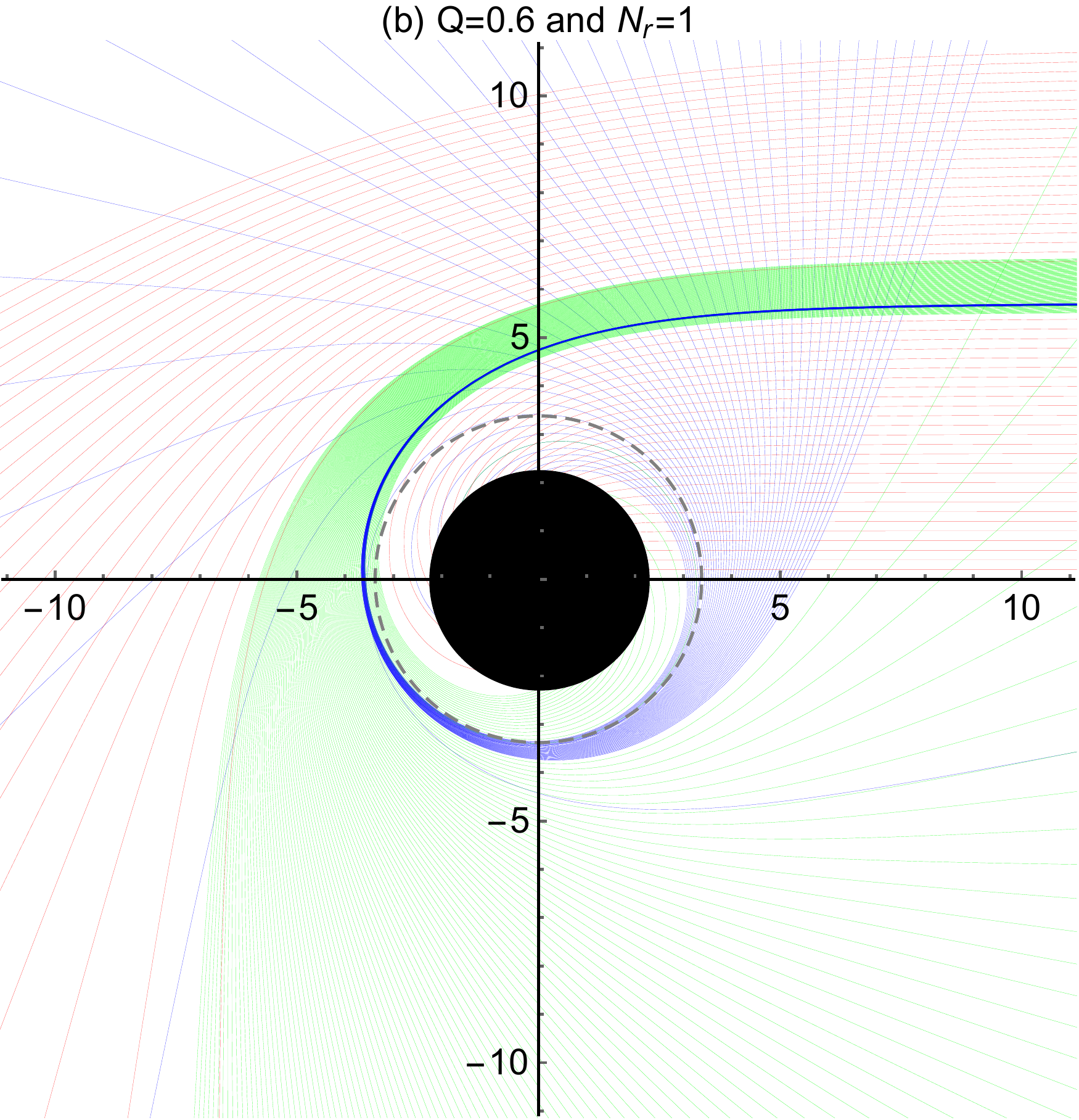}
\hspace{0.2cm}
\includegraphics[width=4.5cm,height=4.5cm]{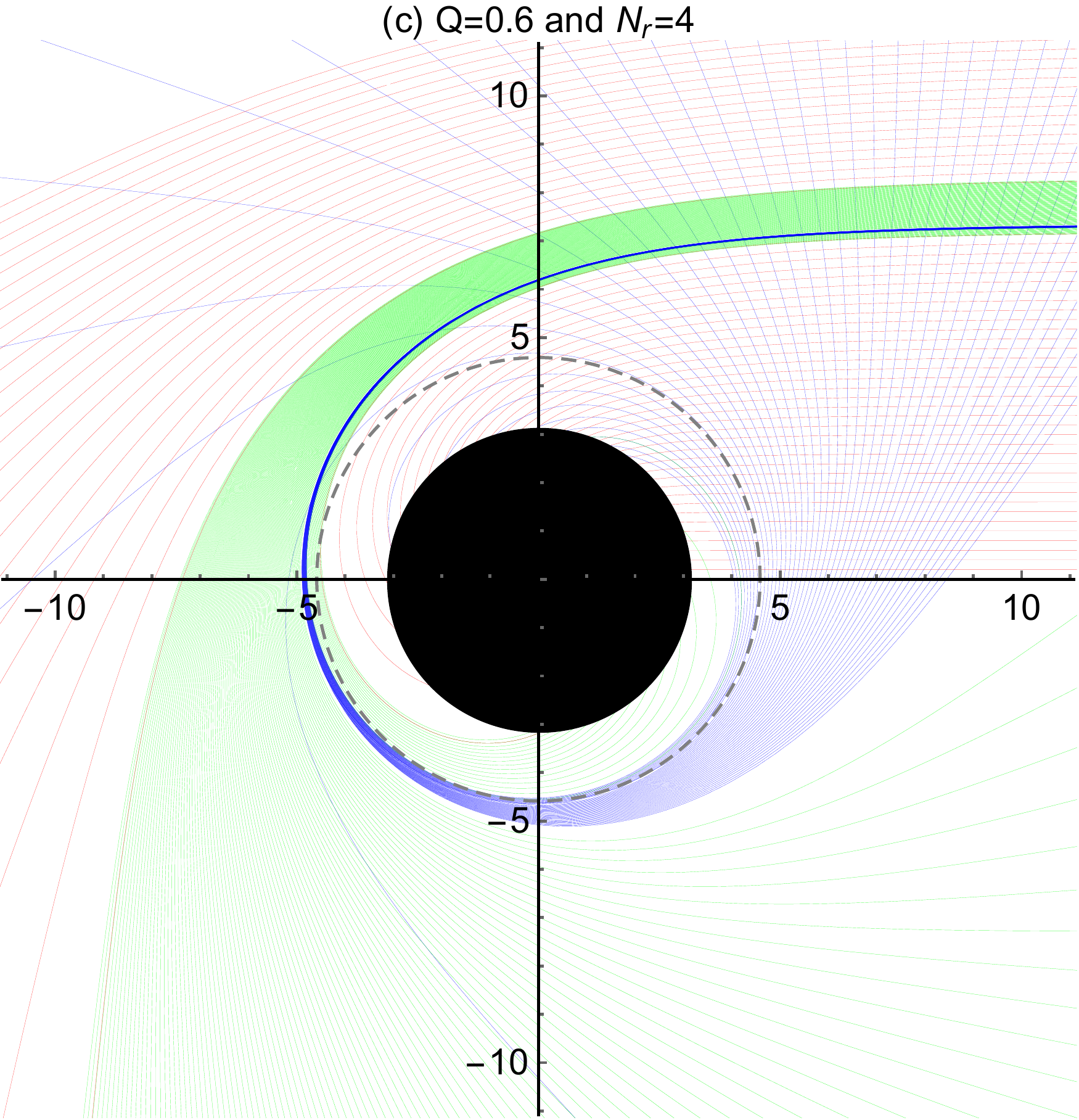}
\hspace{0.2cm}
\includegraphics[width=4.5cm,height=4.5cm]{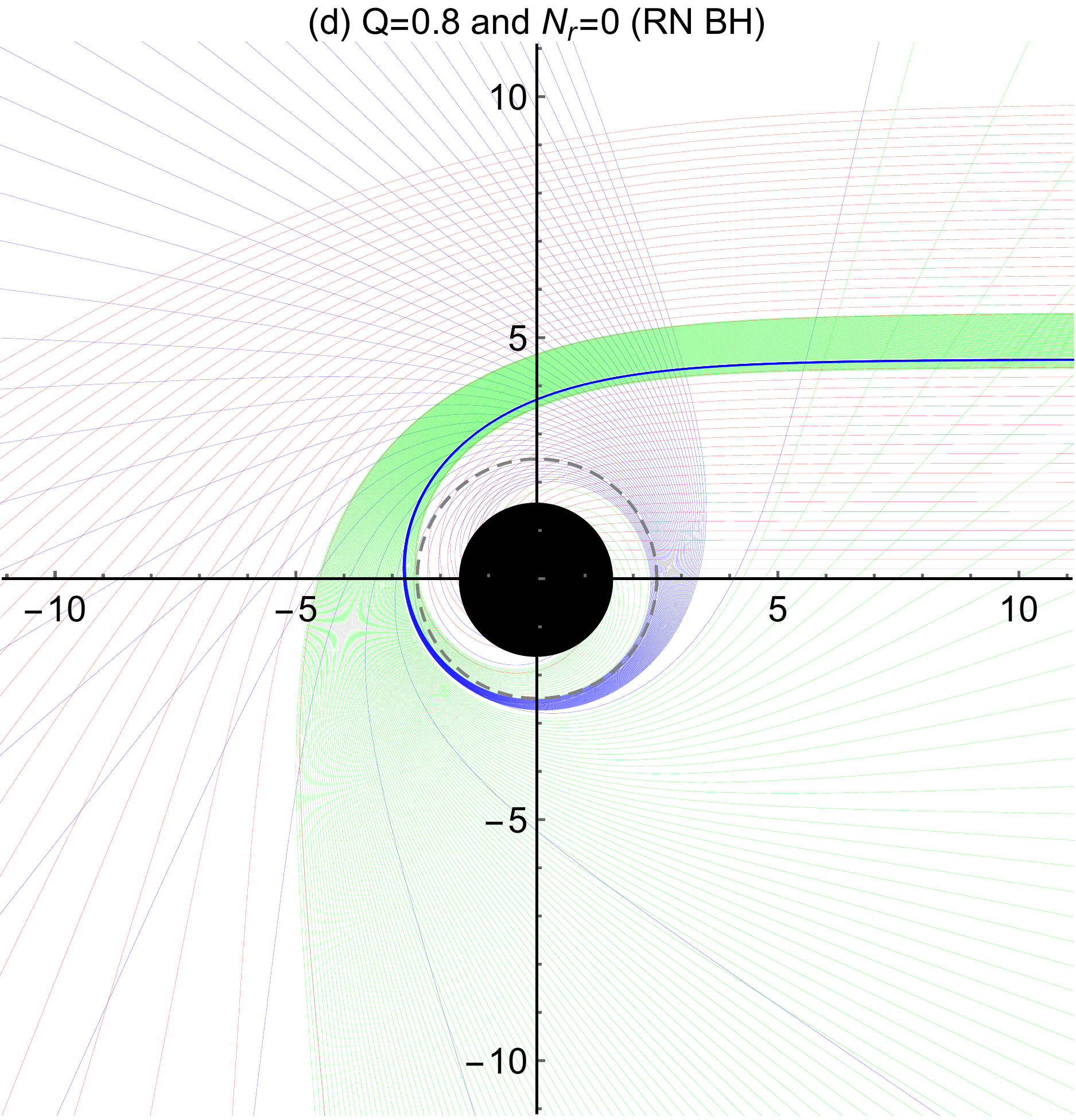}
\hspace{0.2cm}
\includegraphics[width=4.5cm,height=4.5cm]{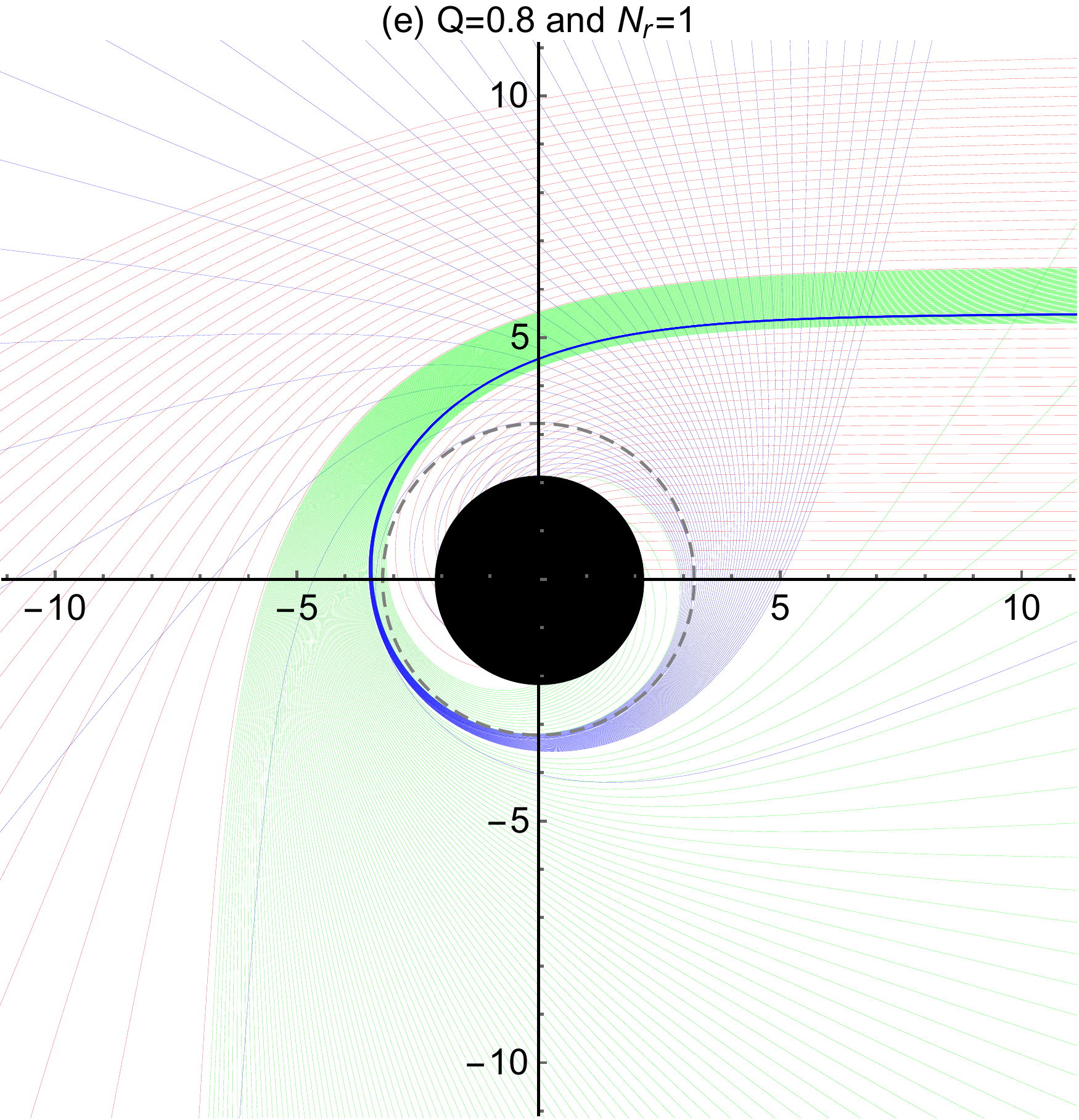}
\hspace{0.2cm}
\includegraphics[width=4.5cm,height=4.5cm]{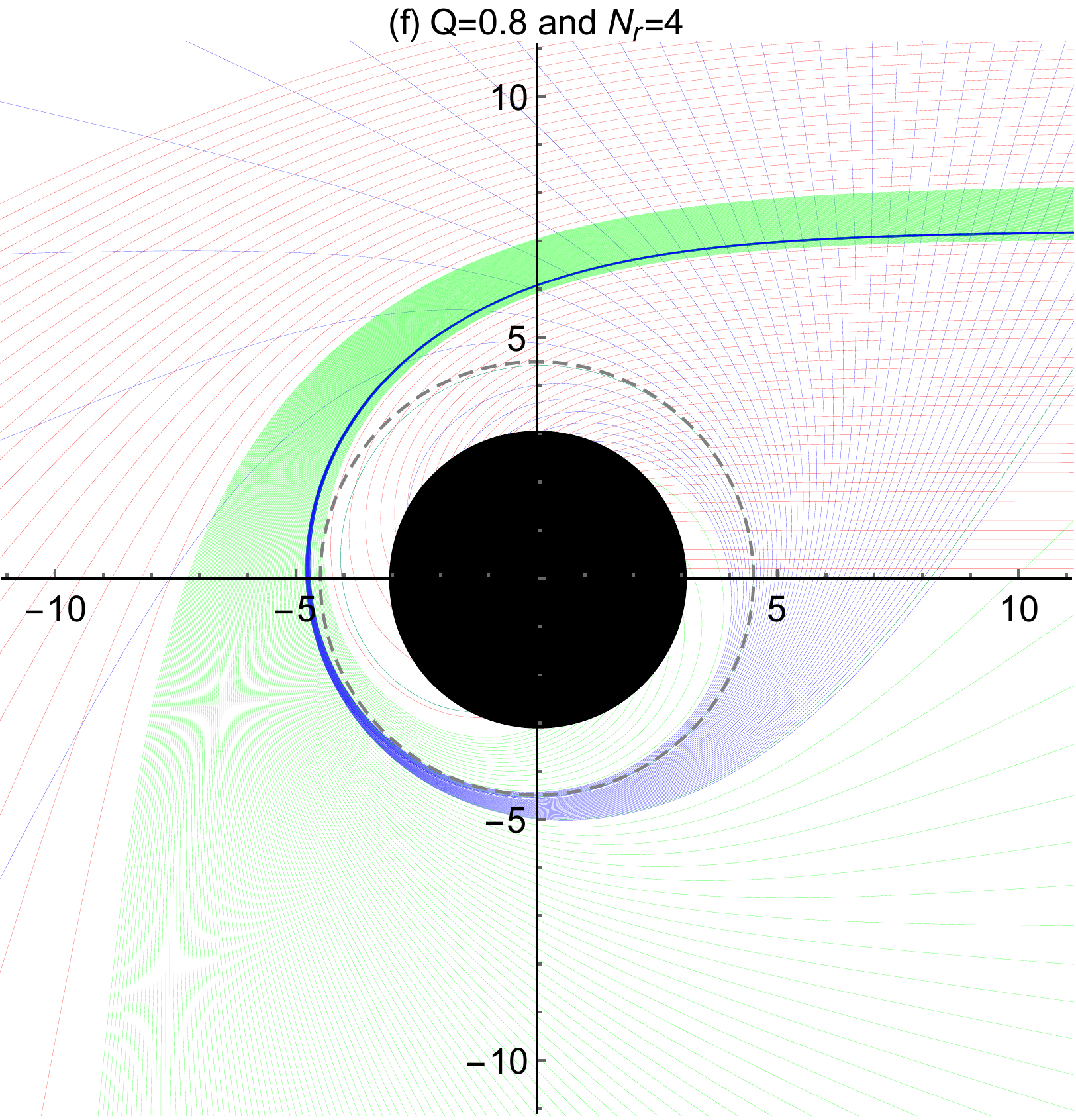}
\parbox[c]{15.0cm}{\footnotesize{\bf Fig~4.}  % figure caption
The light trajectories of different rings in the polar coordinate $(b,\phi)$. The charge BHs in the Rastall gravity are shown as the black disks, and the dashed grey lines represent the BH photon ring orbits. The red lines, green lines, and blue lines correspond to the direct emission, lensing ring, and photon ring, respectively. The BH mass is taken as $M=1$.}
\label{fig4}
\end{center}

\section{Impact of accretion disk radiation position on the BH observational characteristics}
\label{sec:3}
\par
Assuming that the aim BH is surrounded by an optically and geometrically thin accretion disk, we analyze the impact of the disk radiation position on the BH observational characteristics. Considering that the radiation of accretion disk in the universe satisfies gaussian distribution \cite{22}, we parameterize the radiations intensity of the accretion disk as a Gaussian function, i.e.
\begin{equation}
\label{3-1}
I_{\rm emit}(r)~=~\left\{
\begin{array}{rcl}
e^{\frac{-(r-r_{\rm in})^{2}}{8}} ~~~~~~~~~~& & {r>r_{\rm in}},\\
0~~~~~~~~~~~~~~~~~~~~ & & {r\leq r_{\rm in}},\end{array} \right.
\end{equation}
where $r_{\rm in}$ is the innermost radius of the accretion disk. Our analysis is for three different scenarios: (A) $r_{\rm in}=r_{\rm isco}$, where $r_{\rm isco}$ is the innermost stable circular orbit of the BH; (B) $r_{\rm in}=r_{\rm ph}$, where $r_{\rm ph}$ is the radius of the BH photon ring; (C) $r_{\rm in}=r_{+}$, where $r_{+}$ is the radius of the BH event horizon.

\subsection{Case A: $r_{\rm in}=r_{\rm isco}$}
\label{sec:3-1}
\par
The innermost stable circular orbit ($r_{\rm isco}$) is one of the relativistic effects, which represents the boundary between test particles circling and falling into the BH. It can be written as
\begin{equation}
\label{3-1-1}
r_{\rm isco}=\frac{3f(r_{\rm isco})f'(r_{\rm isco})}{2f'(r_{\rm isco})^{2}-f(r_{\rm isco})f''(r_{\rm isco})}.
\end{equation}
We obtain $r_{\rm isco}=6.87 r_{\rm g}$ by taking $Q=0.6$ and $N_{\rm r}=1$. The total radiation intensity $I^{\rm A}_{\rm emit}$ as a function of the radius, the total observed intensity $I^{\rm A}_{\rm obs}$ as a function of the impact parameter, and the two-dimensional image in the celestial coordinates are displayed in the top panels of Fig.5. It is found that the regions of the direct emission, lensing ring, and photon ring are separated. The direct emission starts at $b \simeq 7.84M$ and peaks at $b \simeq 8.28 M$. Its maximum intensity is about $0.49$. The lensing ring is limited to a small range of $b \simeq 5.94M \sim 6.18M$. The photon ring appears at $b \simeq 5.71 M$. In the two-dimensional image, the boundary of the black disk corresponds to $r_{\rm isco}$. A bright lensing ring is shown within the black disk, and the dim photon ring is in the inner of the lensing ring.

\subsection{Case B: $r_{\rm in}=r_{\rm ph}$}
\label{sec:3-2}
\par
Deriving $r_{\rm ph} \simeq 3.38 r_{\rm g}$ ($Q=0.6$ and $N_{\rm r}=1$) from Eq.(\ref{2-5}). We show $I^{\rm B}_{\rm emit}(r)$ as a function of $r$, $I^{\rm B}_{\rm obs}$ as a function of $b$, and the two-dimensional image in the middle panels of Fig.5. Different from Case A, the regions of the direct emission, lensing ring, and photon ring are overlapped. The direct emission starts at $b \simeq 4.29 M$. The very narrow spike at $b \simeq 5.69 M$ is the photon ring, while the broader bump at $b \simeq 5.79 M$ is the lensing ring. The two rings cannot be completely separated. If enlarge the two-dimensional image, one can observe that the photon ring clings to the lensing ring and presents a bright, extremely narrow ring. An utterly dark region is shown in the BH photon ring.
\begin{center}
  \includegraphics[width=5.5cm,height=4cm]{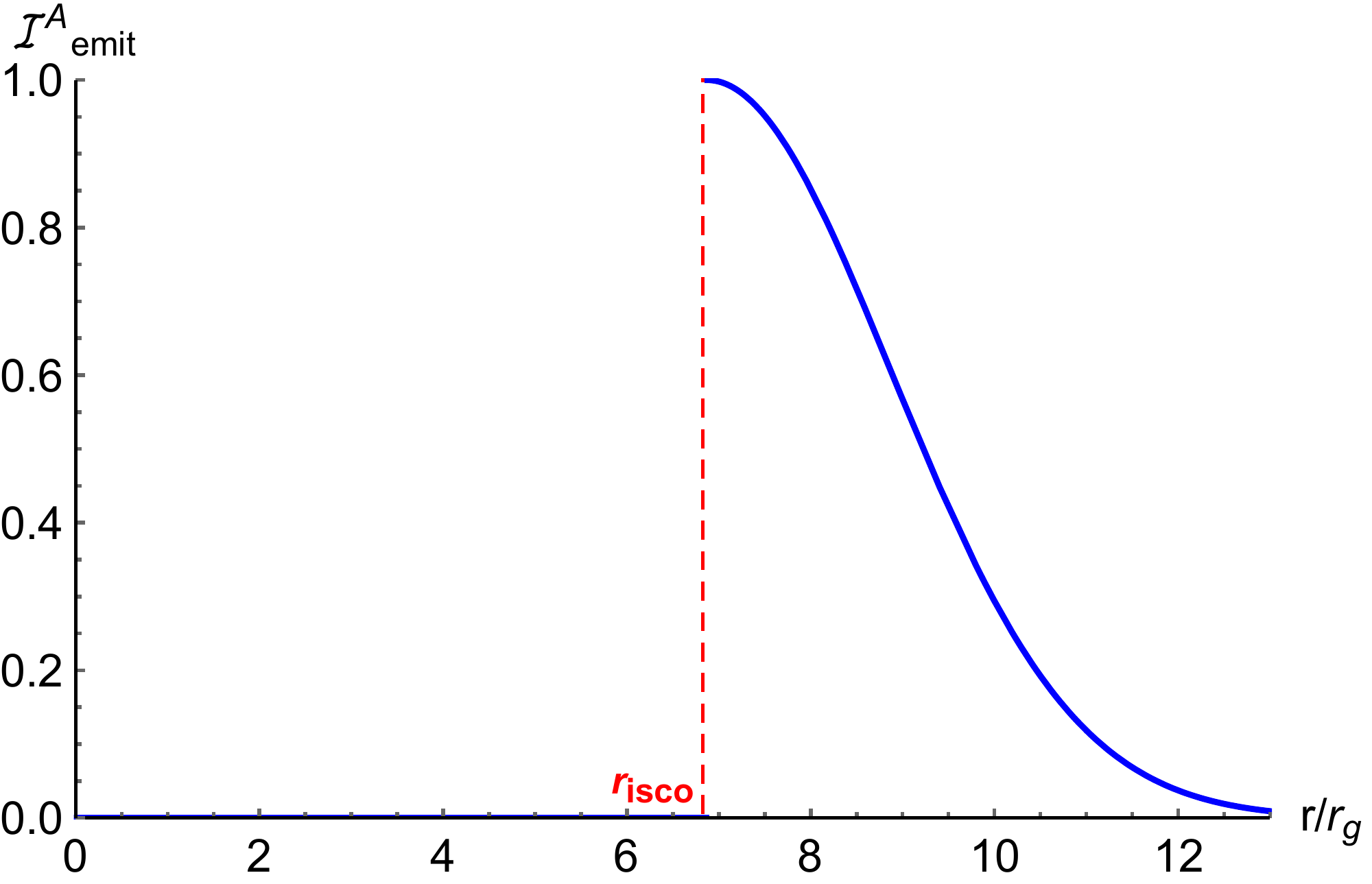}
  \includegraphics[width=5.5cm,height=4cm]{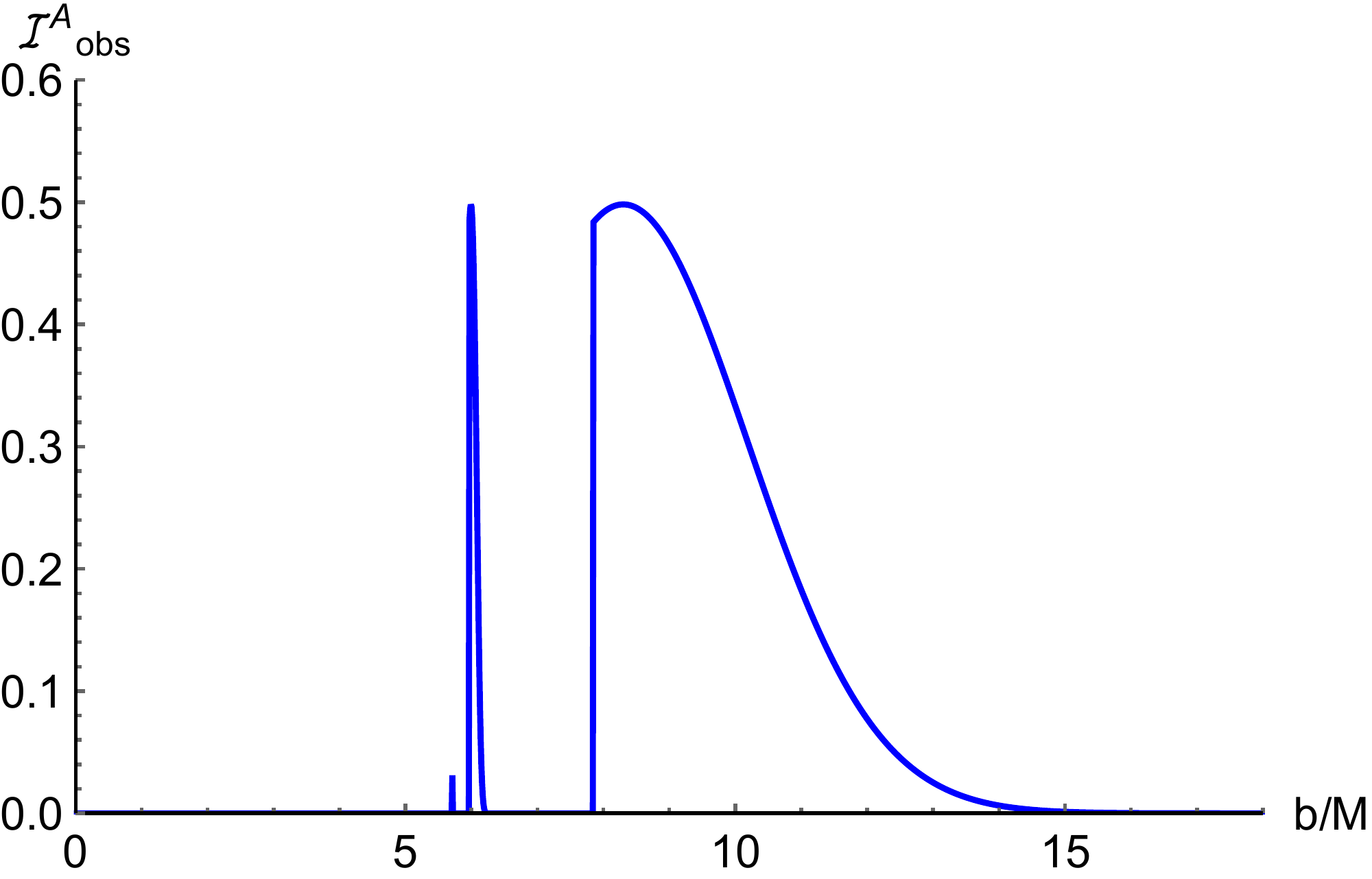}
  \includegraphics[width=4cm,height=4cm]{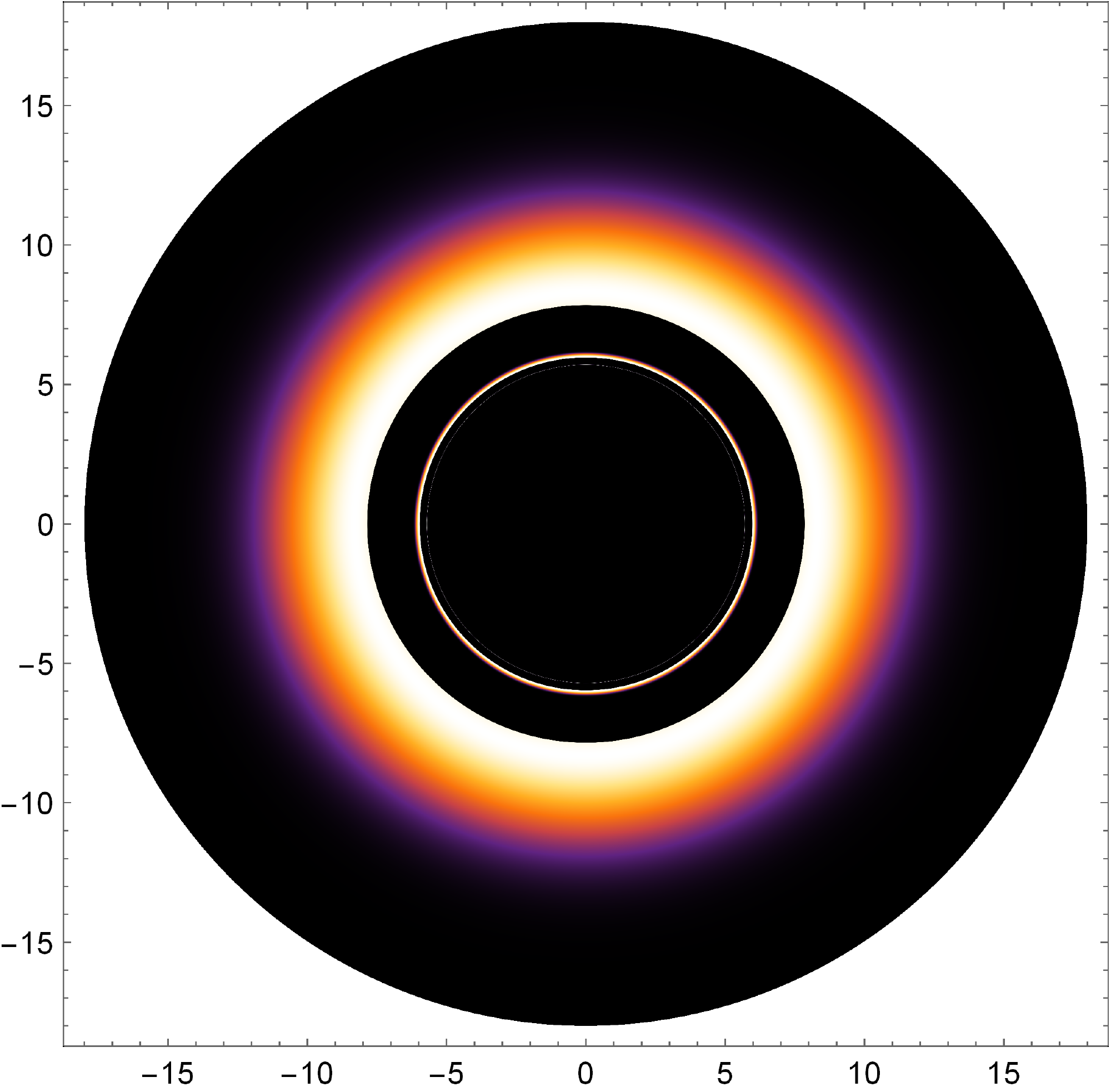}
  \includegraphics[width=5.5cm,height=4cm]{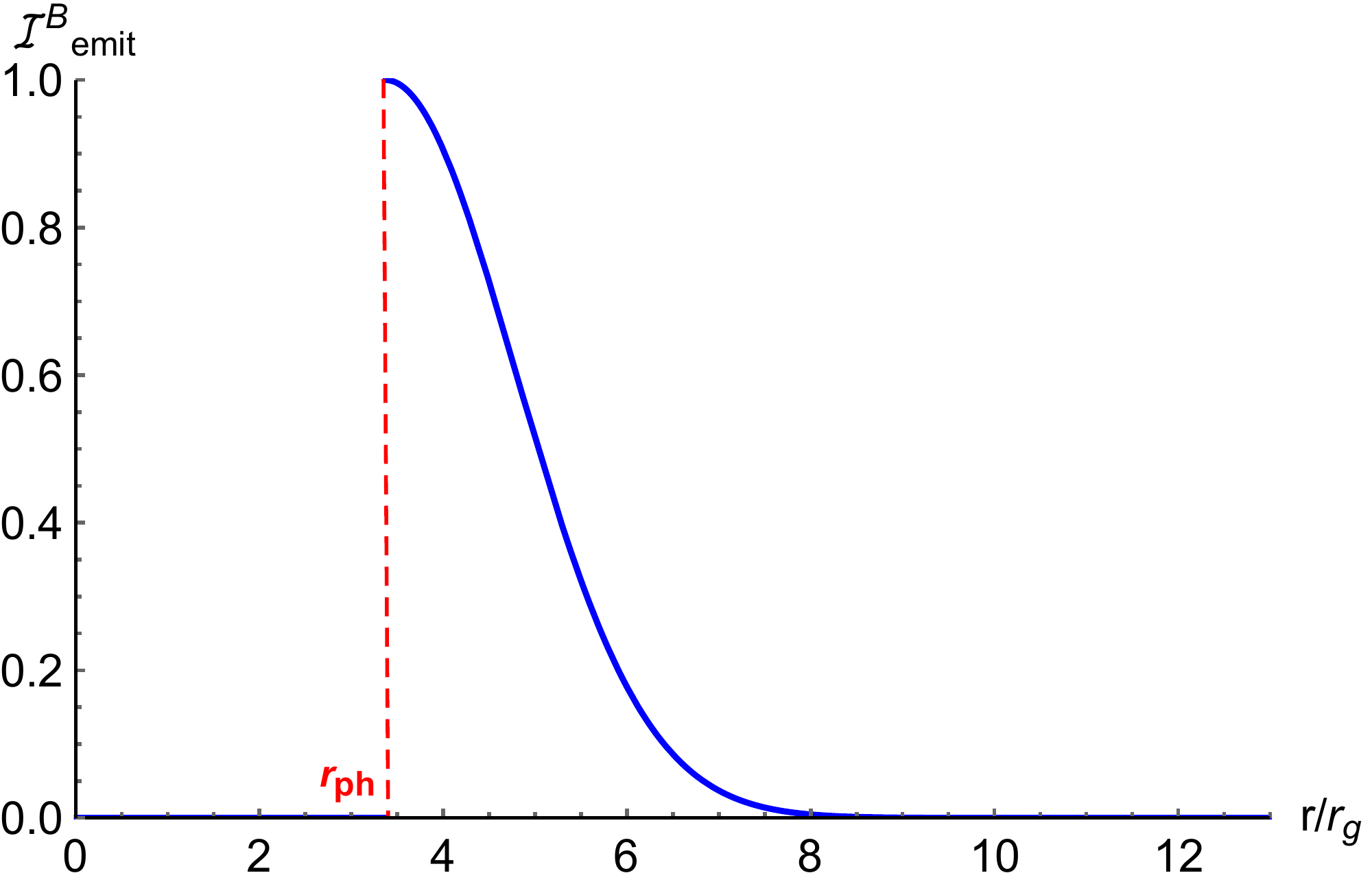}
  \includegraphics[width=5.5cm,height=4cm]{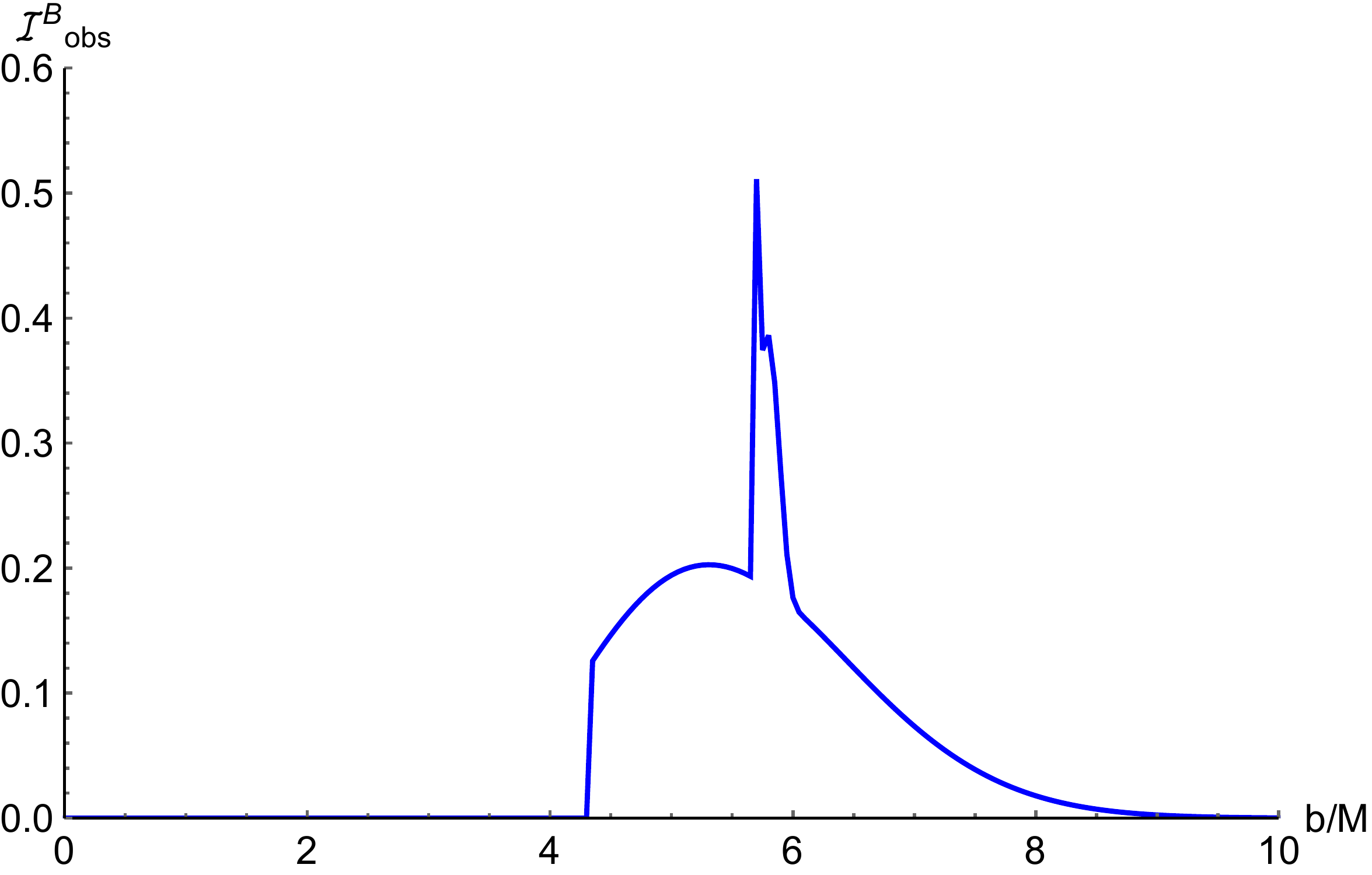}
  \includegraphics[width=4cm,height=4cm]{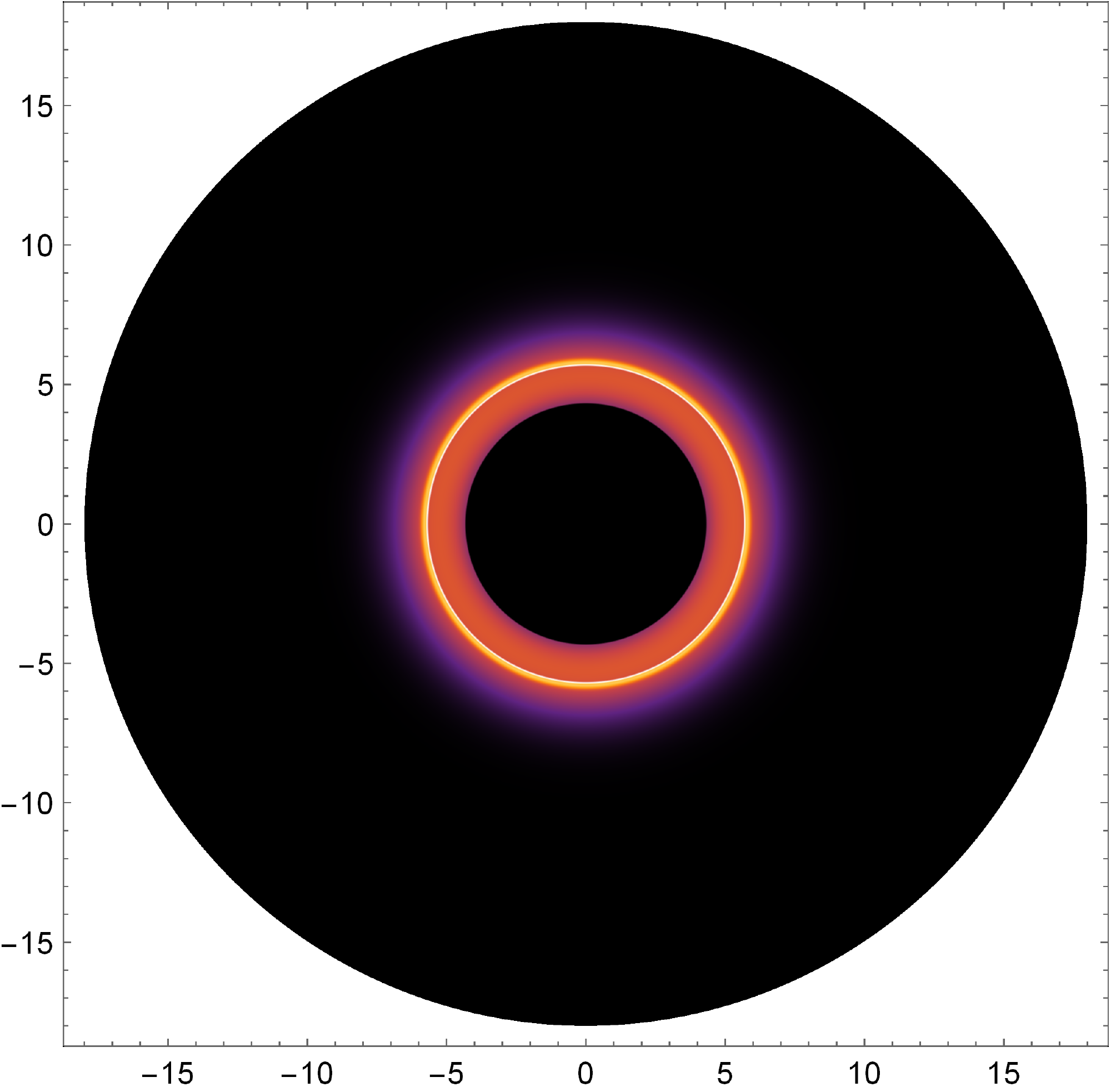}
  \includegraphics[width=5.5cm,height=4cm]{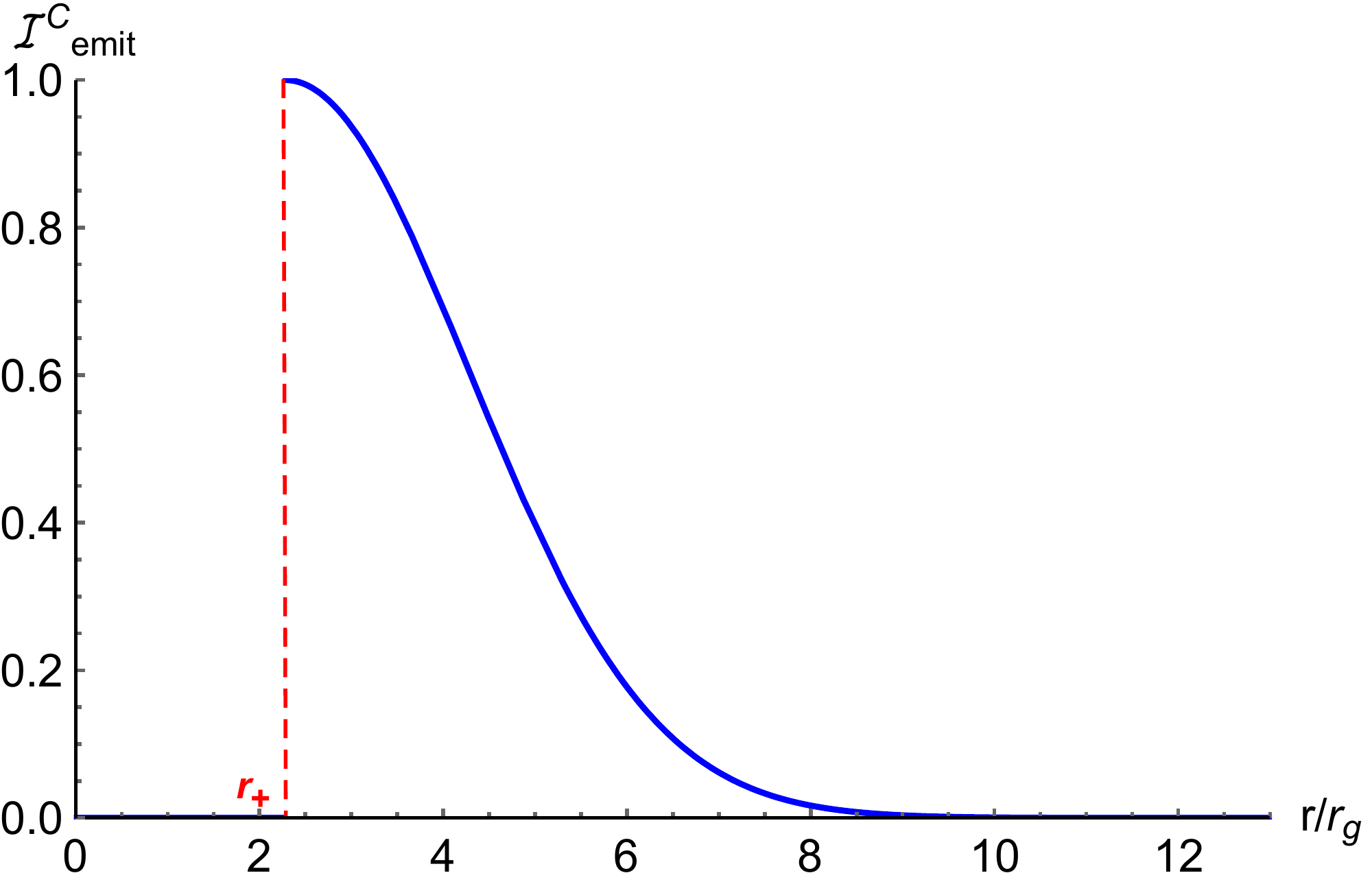}
  \includegraphics[width=5.5cm,height=4cm]{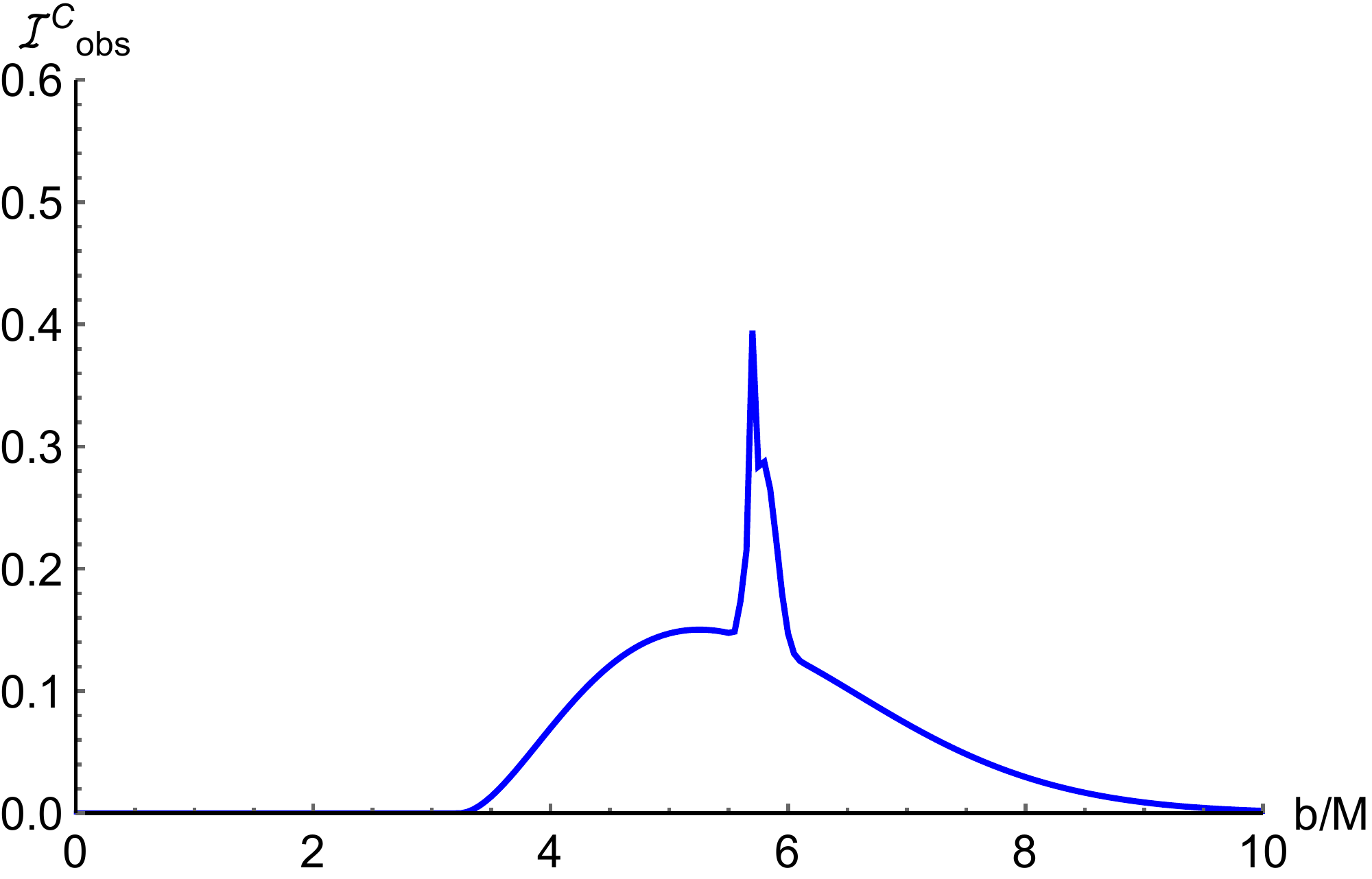}
  \includegraphics[width=4cm,height=4cm]{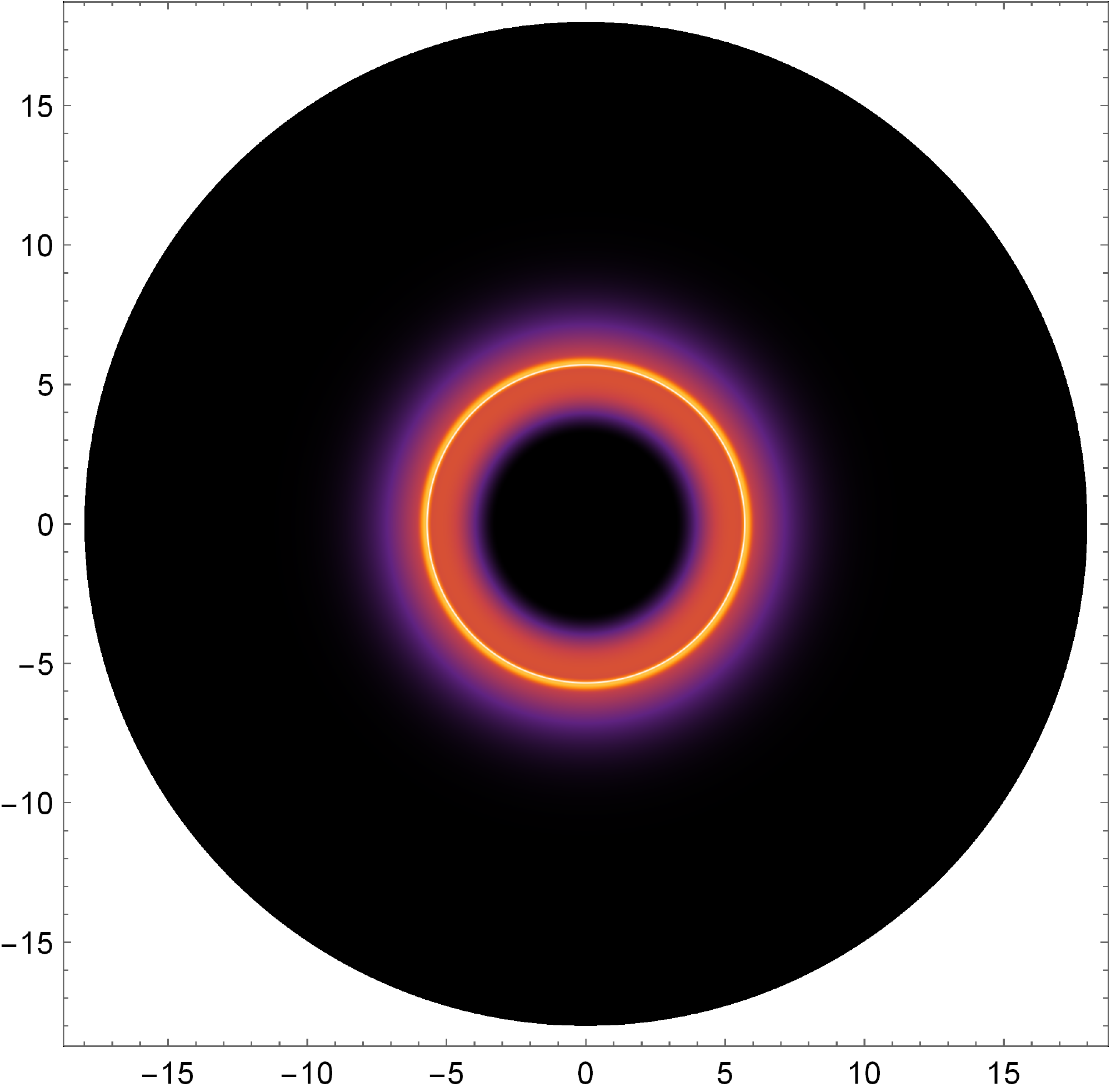}
\parbox[c]{15.0cm}{\footnotesize{\bf Fig~5.}  % figure caption
The total radiation intensity as a function of radius, the total observed intensity as a function of the impact parameter, and the two-dimensional images of the charged BH with PFRF surrounded by the accretion disk in the Rastall gravity. The $Q=0.6$, $N_{\rm r}=1$, and the BH mass is taken as $M=1$.}
\label{fig5}
\end{center}

\subsection{Case C: $r_{\rm in}=r_{\rm +}$}
\label{sec:3-3}
\par
When the innermost of the accretion disk is in the BH event horizon radius ($r_{\rm +}$), the accretion disk radiation peak at the BH event horizon radius $r_{\rm +} \simeq 2.28 r_{\rm g}$ ($Q=0.6$ and $N_{\rm r}=1$). Our results are shown in the bottom panels of Fig.5. This scenario is quite similar to Case B. It is found that the black area that can be observed is reduced to the BH event horizon, but the position and shape of the lensing ring and photon ring change very weakly. The results suggest that the observable characteristics of the charged BH shadow in the Rastall gravity depend on the position of the accretion disk, the feature of the rings depend on the BH itself.

\par
Our above analysis is for $N_{\rm r}>0$. Note that in the limit of $N_{\rm r} \rightarrow 0$, the BH metric (\ref{2-2}) returns to the usual RN BH. We compare the images between the RN BH and the charged BH surrounded by a PFRF in the Rastall gravity by adopting $N_{\rm r}=4$ in Fig.6 for the case (A). One can observe that a bright lensing ring is shown within the innermost stable circular orbit, and the dim photon ring is in the inner of the lensing ring for $N_{\rm r}=0$ (the RN BH). Differently, the photon ring disappears and the lensing ring is displayed as an extremely narrow ring in case of $N_{\rm r}=4$. This would be an effective characteristic to distinguish the charged BH in the Rastall gravity from the RN BH in the Einstein gravity. Observation of such a feature depends on the EHT resolution.
\begin{center}
  \includegraphics[width=4.5cm,height=4cm]{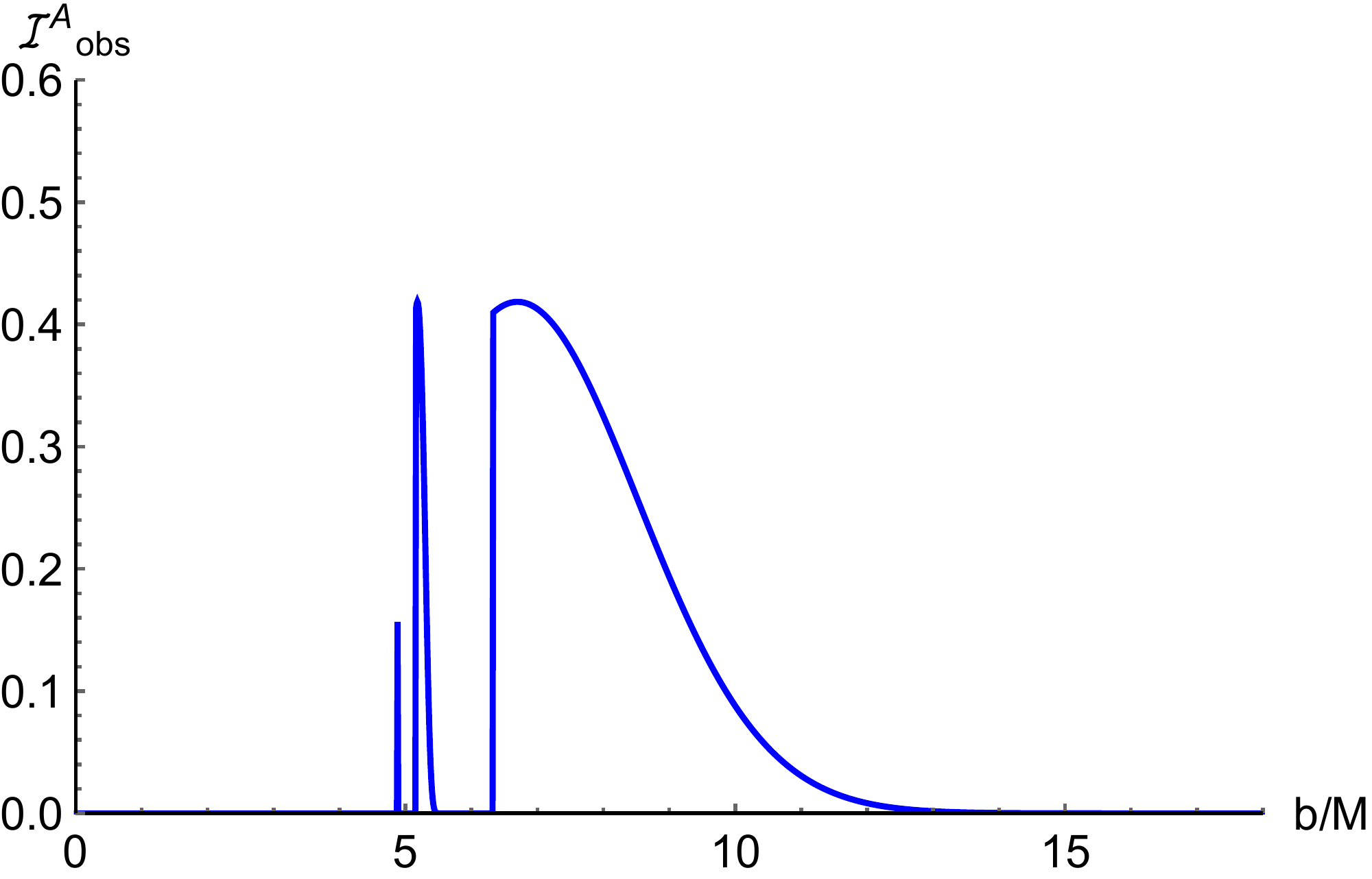}
  \hspace{0.7cm}
  \includegraphics[width=4cm,height=4cm]{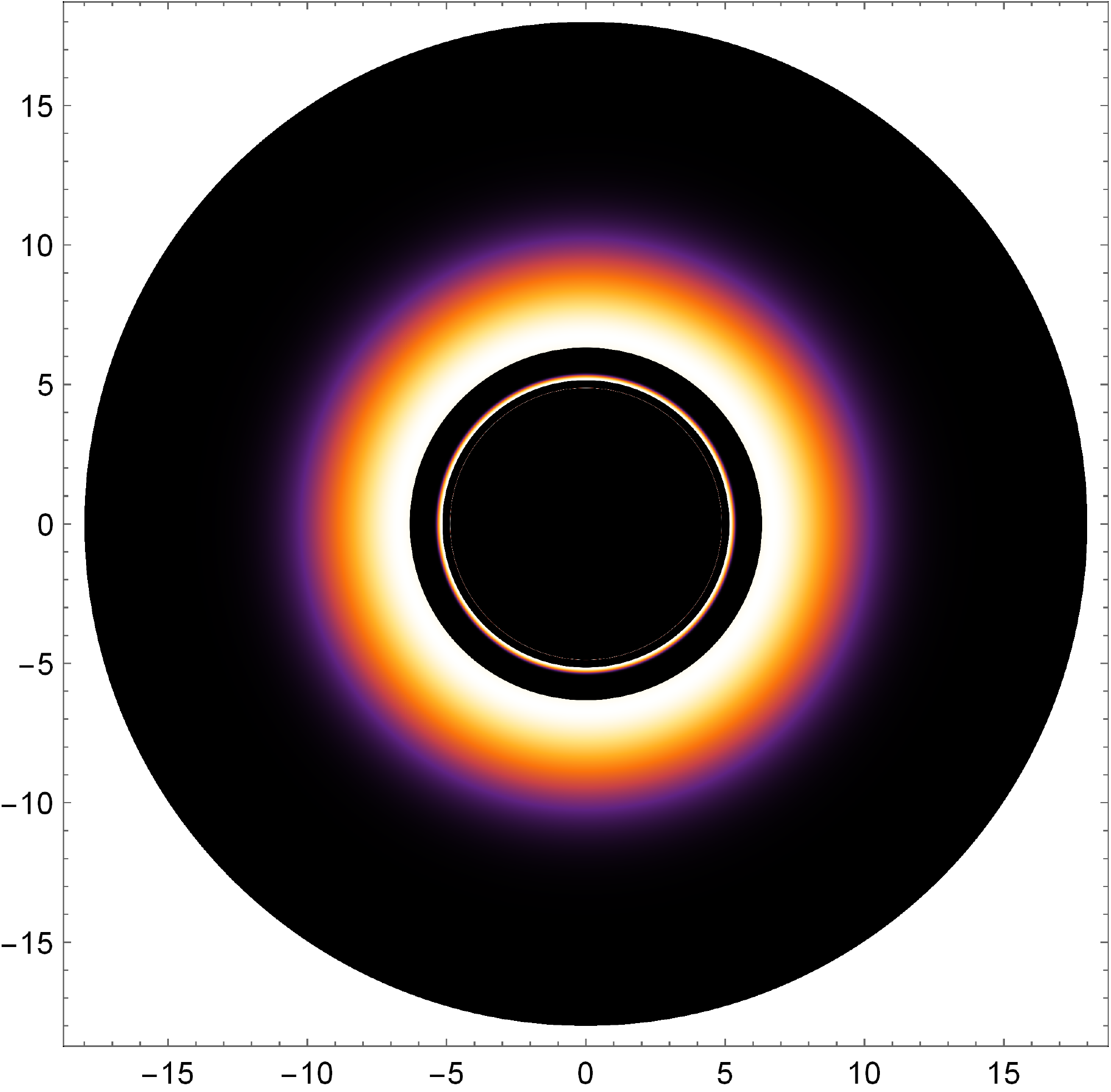}
  \hspace{0.7cm}
  \includegraphics[width=4cm,height=4cm]{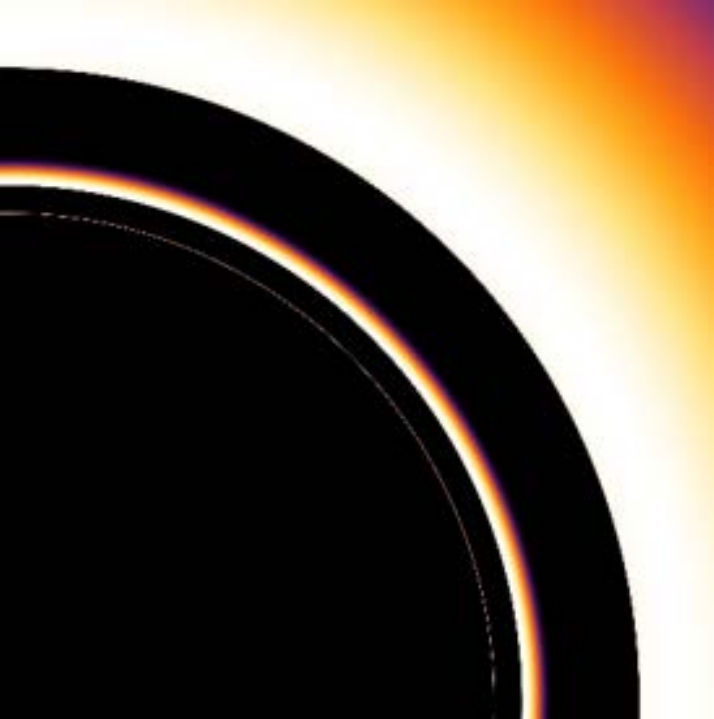}
  \hspace{0.7cm}
  \includegraphics[width=4.5cm,height=4cm]{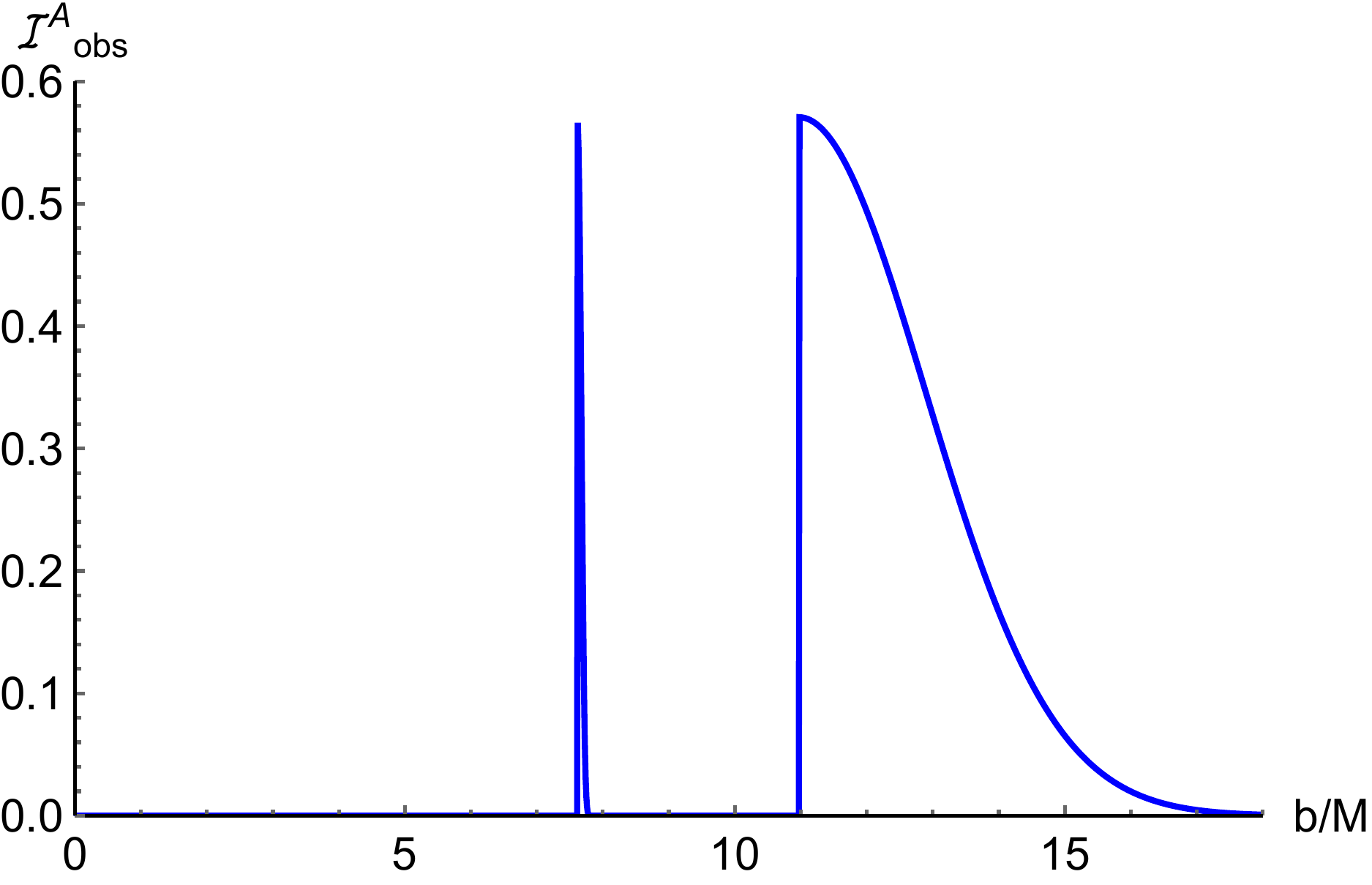}
  \hspace{0.7cm}
  \includegraphics[width=4cm,height=4cm]{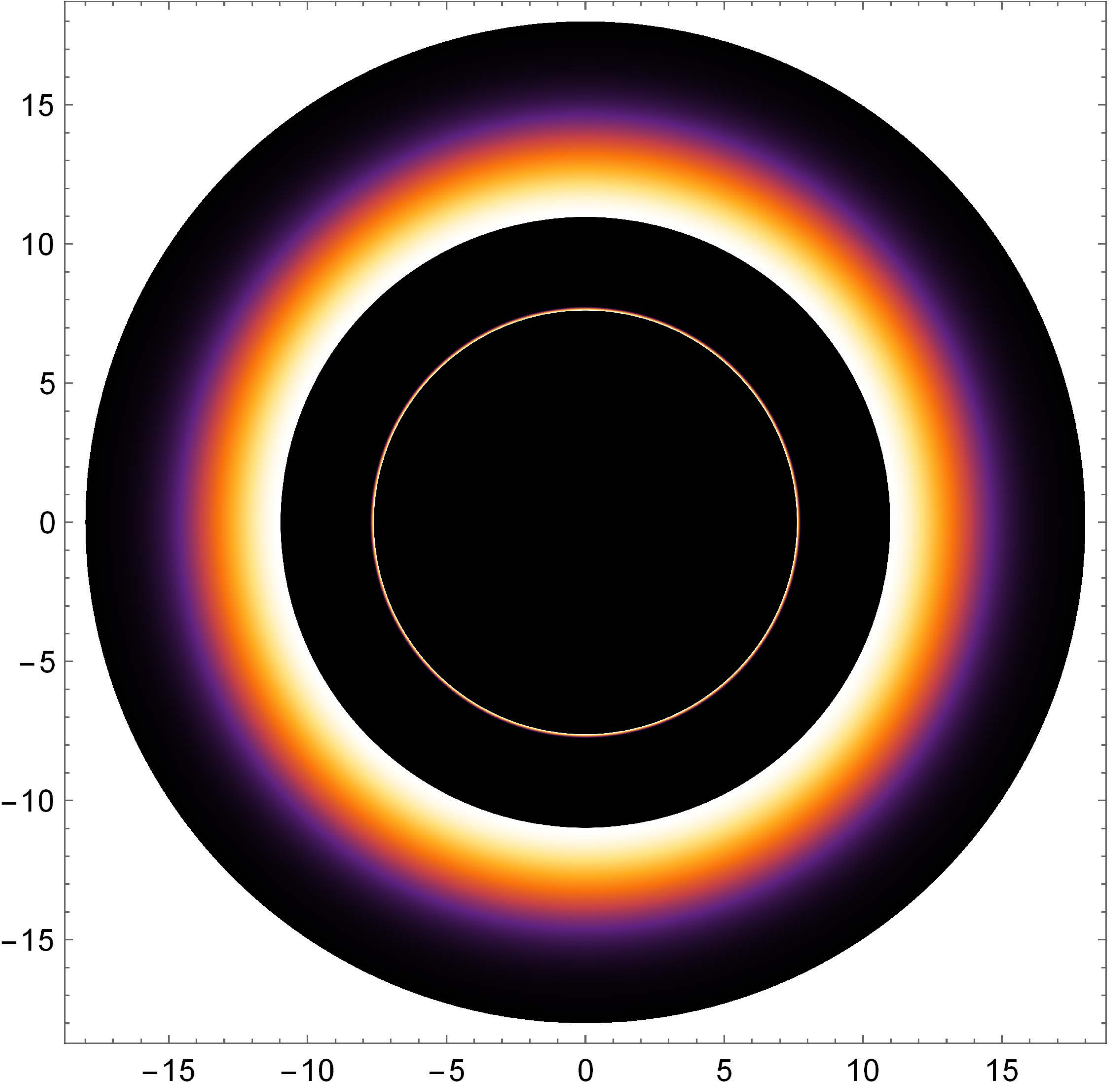}
  \hspace{0.7cm}
  \includegraphics[width=4cm,height=4cm]{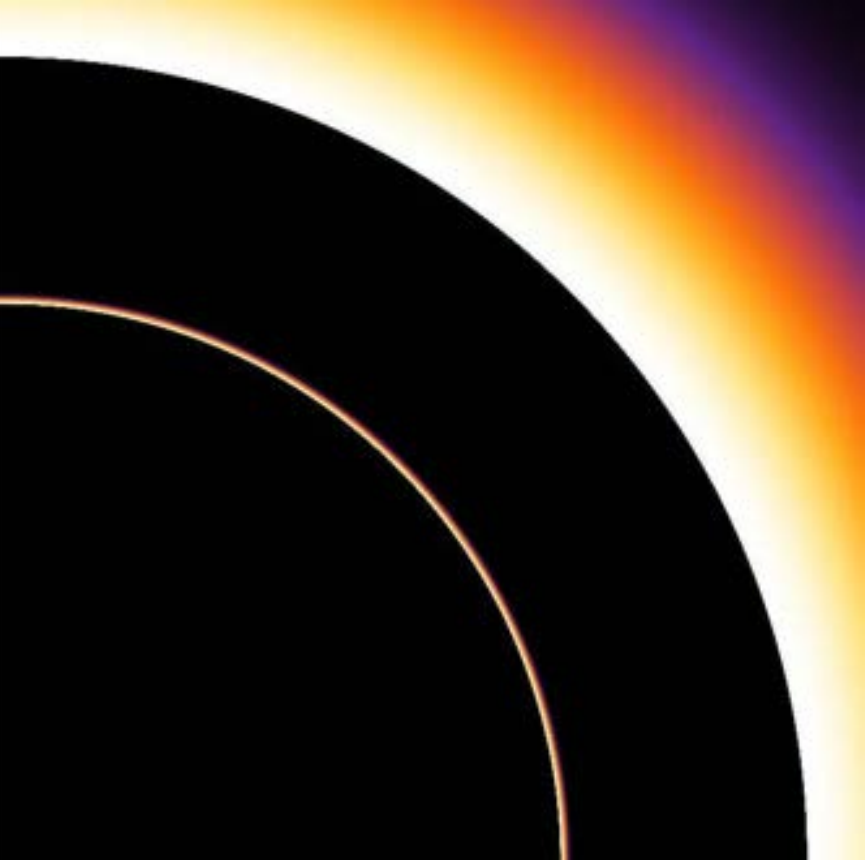}
\parbox[c]{15.0cm}{\footnotesize{\bf Fig~6.}  % figure caption
The total observed intensity as a function of the impact parameter ({\em left panels}) for $Q=0.6$ and the corresponding two-dimensional images ({\em middle panels}) together with zooming in part of the image for illustrating their lensing and photon rings ({\em right panels}). The top and bottom panels are for $N_{\rm r}=0$ (for the RN BH) and $N_{\rm r}=4$, respectively. The BH mass is taken as $M=1$.}
\label{fig6}
\end{center}

\par
We blur the two-dimensional image and make them correspond roughly to the EHT resolution, as shown in Fig.7. The simple blur does not correspond to the EHT image reconstruction and can only offer a rough illustration of the EHT resolution. From the bottom panels of Fig.7, one can observe that the blurring washes out the lensing ring and photon ring features. Their observational appearances reply on instrument resolution. It is difficult to obtain the ring information with the current resolution of EHT. The top panels of Fig.7 correspond to the M87$^{*}$ image, and the shadow of a charged BH with a PFRF under the static and infalling spherical accretions context in the Rastall gravity. We can see that although the observed shadow luminosities are different, the size of the BH shadow does not change, which means that the BH shadow is shown as a geometric feature of space-time in the scenario of the spherical accretions.
\begin{center}
  \includegraphics[width=4.2cm,height=4.2cm]{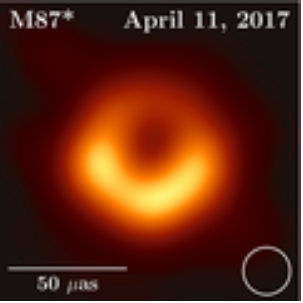}
  \includegraphics[width=4.2cm,height=4.2cm]{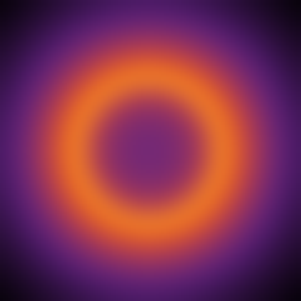}
  \includegraphics[width=4.2cm,height=4.2cm]{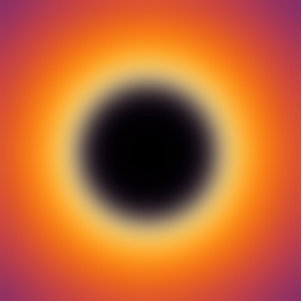}
  \includegraphics[width=4.2cm,height=4.2cm]{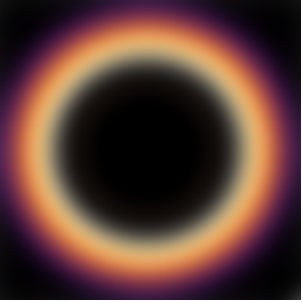}
  \includegraphics[width=4.2cm,height=4.2cm]{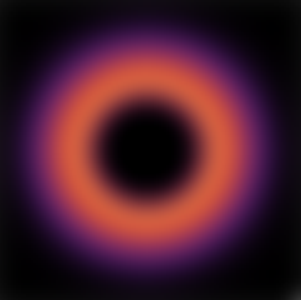}
  \includegraphics[width=4.2cm,height=4.2cm]{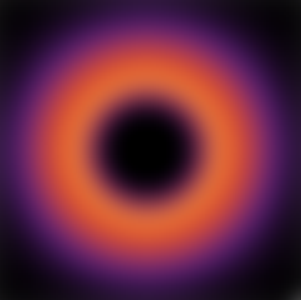}
\parbox[c]{15.0cm}{\footnotesize{\bf Fig~7.}
Blurred two-dimensional images utilizing a Gaussian filter with a standard deviation of 1/12 the field of view.}
\label{fig7}
\end{center}

\section{Conclusions and Discussion}
\label{sec:4}
\par
We have investigated the observable characteristics of the charged BH with PFRF surrounded by a thin disk accretion in the Rastall gravity. We found that the increase of the BH charge leads to the decrease of the the BH event horizon radii, shadow radii, and critical impact parameters, displaying that the BH photon ring is shrunk inward the BH by increasing the charge. While the radiation field parameter has the opposite effect. It is also found that the radii of the direct emission, lensing ring, and photon ring are dramatically increased as the radiation field parameter increases, but they only weakly depend on the BH charge. The light rays are more curved near the BH for a larger BH charge since the increase of the BH charge leads to increased space-time curvature.

\par
Assuming that a thin accretion disk surrounds the BH, we showed the observational characteristics of the shadows and rings in three cases. In case the innermost radius of the accretion disk equals $r_{\rm isco}$, we found that the regions of the direct emission, lensing ring, and photon ring are separated. A bright lensing ring is shown within the $r_{\rm isco}$, and the dim photon ring is in the inner of the lensing ring. If the innermost radius of the accretion disk equals to $r_{\rm ph}$, the lensing and photon rings are embedded in the direct emission region, and these two rings cannot be completely separated. In case the innermost radius extends to the event horizon of the BH ($r_{\rm +}$), the observable black area is shrunk to the BH event horizon, but the position and shape of the lensing ring and photon ring change very weakly. The total observed flux of the charged BH in the Rastall gravity is dominated by the direct emission, the lensing ring provides a small contribution, and the photon ring is negligible. These results suggest that the observable characteristics of the charged BH surrounded by the thin disk accretion in the Rastall gravity depend on both the BH space-time structure and the position of the radiating accretion disk with respect to the BH. As shown in \cite{21}, the size of the BH shadow does not change in the spherical accretion scenario, but it relies on the position of the radiating accretion disk with respect to the BH in the Rastall gravity. Additionally, a ring feature appears in this analysis, which can be used as an important feature to distinguish spherical accretion and disk accretion models. However, this feature still cannot be observed with the current resolution capacity of the EHT.

\par
The charges of the some modified gravity BHs are stringently constrained by using the M87$^{*}$ observations. In the framework of the standard Einstein gravity, the charge of the RN BH is constrained as $Q \leq 0.9$ within $1 \sigma$ confidence level, implying that the BH of M87$^{*}$ is not a highly charged dilaton BH \cite{Koch}. Similarly, we place constraints the Rastall gravity parameters and verify the images of the charged BH in the Rastall gravity with observations of the BH image of M87$^{*}$. We obtained that $Q \leq 1.03$ and $N_{\rm r} \leq 3.23$ at $1 \sigma$ confidence level, and $Q \leq 1.55$ and $N_{\rm r} \leq 6.39$ at $2 \sigma$ confidence level for the charged BH surrounded by PFRF in the Rastall gravity. Meanwhile, we compare the images between the RN BH and the charged BH surrounded by a PFRF in the Rastall gravity. It is found that a bright lensing ring is shown within the innermost stable circular orbit, and the dim photon ring is in the inner of the lensing ring for the RN BH. While the photon ring disappears and the lensing ring is displayed as an extremely narrow ring in case of $N_{\rm r}=4$. This would be an effective characteristic to distinguish the charged BH in the Rastall gravity from the RN BH in the Einstein gravity.

\section*{Acknowledgments}
This work is supported by the National Natural Science Foundation of China (Grant No. 12133003, 11851304, and U1731239), by the Guangxi Science Foundation and special funding for Guangxi distinguished professors (2017AD22006).

\clearpage

\end{CJK}
\end{document}